\documentclass[useAMS,usenatbib]{mn2e}
\usepackage{aas_macros}
\usepackage[a4paper,centering, totalwidth=520pt, totalheight=700pt]{geometry}
\def\k#1 {k_{{\rm #1}}}

\def\mH2p{H_2^+}

\def\ltsima{$\; \buildrel < \over \sim \;$}
\def\simlt{\lower.5ex\hbox{\ltsima}}   
\def\gtsima{$\; \buildrel > \over \sim \;$}
\def\gtsim{\lower.5ex\hbox{\gtsima}}

\usepackage{graphicx}
\usepackage{amssymb}
\usepackage{amsmath}
\usepackage{enumerate}

\usepackage[usenames,dvipsnames]{color}

\title[Simulating collisionless DM fluids]{A new approach to simulating collisionless dark matter fluids}
\author[O. Hahn, T. Abel \& R. Kaehler]{Oliver Hahn\thanks{Email:
    hahn@phys.ethz.ch}$^{1,2}$, Tom Abel$^1$\thanks{Email:
    tabel@stanford.edu} \& Ralf Kaehler\thanks{Email:
    kaehler@slac.stanford.edu} $^1$ \\
  $^{1}$Kavli Institute for Particle Astrophysics and Cosmology, \\ 
  \quad Stanford University, SLAC National Accelerator Laboratory, Menlo Park, CA 94025, USA\\ 
  $^{2}$Department of Physics, ETH Zurich, CH-8093 Z\"urich, Switzerland
  }

\begin{document}

\date{submitted to MNRAS}
\pagerange{\pageref{firstpage}--\pageref{lastpage}} \pubyear{2013}
\maketitle

\label{firstpage}

\begin{abstract}
Recently, we have shown how current cosmological $N$--body codes already follow the fine grained phase--space information of the dark matter fluid. Using a tetrahedral tesselation of the three-dimensional manifold that describes perfectly cold fluids in six-dimensional phase space, the phase-space distribution function can be followed throughout the simulation. This allows one to project the distribution function into configuration space to obtain highly accurate densities, velocities, and velocity dispersions. Here, we exploit this technique to show first steps on how to devise an improved particle--mesh technique. At its heart, the new method thus relies on a piecewise linear approximation of the phase space distribution function rather than the usual particle discretisation. We use pseudo-particles that approximate the masses of the tetrahedral cells up to quadrupolar order as the locations for cloud--in--cell (CIC) deposit instead of the particle locations themselves as in standard CIC deposit. We demonstrate that this modification already gives much improved stability and more accurate dynamics of the collisionless dark matter fluid at high force and low mass resolution. We demonstrate the validity and advantages of this method with various test problems as well as hot/warm--dark matter simulations which have been known to exhibit artificial fragmentation. This completely unphysical behaviour is much reduced in the new approach. The current limitations of our approach are discussed in detail and future improvements are outlined. 
\end{abstract}

\begin{keywords}
cosmology: theory, dark matter, large-scale structure of Universe -- galaxies: formation -- methods: N-body, numerical
\end{keywords}

\maketitle

\section{Introduction}
In the current cosmological standard model, structure formation is dominated by the 
gravitational collapse of a collisionless fluid of cold dark matter (CDM). $N$-body simulations
have been an immensely successful tool to study the gravitational collapse of tiny density
perturbations into dark matter haloes that grow hierarchically by merging to assemble
ever more massive systems 
\citep[e.g.][and references therein]{Davis:1985, Efstathiou:1985, Bertschinger:1998,Springel:2005a}. 
In cosmological $N$-body simulations, the initial 
density distribution -- from which all structure forms through
gravitational instability -- is sampled with a finite number of particles,
many orders of magnitude fewer than the actual number of microscopic dark matter particles
contained in the system. Ideally, the system would be treated in the continuum limit as a collisionless
fluid. However, sampling the distribution function with a finite number of particles introduces
interactions from few-particle terms that would be absent in the fluid
limit.  This manifests itself in two-body relaxation effects, particularly
due to close encounters of particles whose mutual gravitational interaction is unbounded
if the particles were true point particles. Clearly the force between any
two particles has to be limited and this is typically achieved by
softening the interaction potential on some physical scale. This
introduces the force resolution as a parameter in the calculation
which in general is  related to the global mass resolution of the system \citep[cf. e.g.][]{Power:2003}. Thus, the aim to
resolve the inner parts of haloes with as few particles as possible has led to a consensus
in the field that a force resolution of $1/30$--$1/60$ of the mean linear particle separation
ought to be used as a compromise between keeping collisionality low and having a 
high force resolution.

However, it has long been recognized that when force and mass resolution of cosmological 
$N$--body simulations are not matched,  rather unphysical results can be obtained \citep[e.g.][]{Centrella:1983,
  Centrella:1988, Peebles:1989, Melott:1989, Diemand:2004, Melott:1997, Splinter:1998, Wang:2007a, Melott:2007, Marcos:2008, Bagla:2009}.
  This is particularly apparent in regions of modest overdensity where high force resolution acts
  on a low mean particle count in a convergent flow.
Some effort has thus been invested into the development of simulation techniques which may
overcome these obvious shortcomings. All the work in three dimensions has been focused on 
devising adaptive force softening algorithms that try to minimize unphysical two--body effects. 
In these approaches, the force resolution is varied to match the local mass resolution~\citep{Knebe:2000, Price:2007, Bagla:2009,
Iannuzzi:2011}. These approaches provide more physical solutions in a number of idealized test problems, such as e.g. the
plane wave collapse of~\cite{Melott:1997}, compared to the standard approaches.

At the same time, when applying these codes to typical cosmological applications, solutions are 
obtained that can differ significantly from those using the standard techniques in various aspects. \cite{Knebe:2000}, e.g., give a case in
which the central density of their simulated cold dark matter haloes is more than a factor of two higher in the simulations with adaptive
force softening.  \cite{Iannuzzi:2011} show that with gravitational softening that is adaptive in space and time, small
scale noise still is not entirely suppressed. The reason for this is an attractive force appearing in the conservative
formulation that can be analytically derived from the equations
of motion with the symmetrized conservative adaptive softening term (A.~Hobbs \& J.~Read, private communication).
Hence, this does not alleviate the upturn in the halo mass function at the low mass end (F.~Iannuzzi, private communcation),
quite in accord with the hot/warm dark matter case shown, e.g., in 
\cite{Wang:2007a} (their Figure~9, but see also our discussion below). 
 
It is still quite unclear whether the known shortcomings of the N-body method with high force and low mass resolution
have a major impact on the results of CDM simulations. The main problem to bring these loose ends together
lies in the nature of CDM structure formation: small objects form first -- and the first physical objects always on
scales that are not resolved in the simulation --
and larger objects form by hierarchical growth. The formation of structure in CDM is thus inherently granular,
so that the additional granularity of the $N$-body method might possibly be only a subordinate effect
in simulations of CDM structure formation. However, this is an unverified assumption and
still remains to be demonstrated. \cite{Ludlow:2011}, e.g., find a significant fraction of low mass haloes in 
CDM simulations that cannot be matched to peaks in the initial conditions and are arguably a pure
artefact of the $N$--body method.

In contrast to CDM, it has been realized many times and appears to be general consensus by now
that the $N$-body method works significantly
less well in warm dark matter (WDM) or hot dark matter (HDM) simulations. In these simulations, the initial power spectrum,
from which the density perturbations for the simulations are sampled, has exponentially
suppressed power on scales below a free-streaming scale that is directly related to the
rest mass of the respective dark matter particle candidate. The fact that for dark matter particles
with finite mass also a finite velocity distribution function exists at any point in space
is typically ignored on the grounds that the width of this distribution can be neglected 
with respect to the velocities arising from gravitational collapse. The fluid is thus treated
in the perfectly cold limit during the non-linear evolution. These HDM/WDM simulations then employ
the $N$-body method to evolve a system of particles obeying the density and velocity
power spectra with small scale suppression~\citep{Bode:2001, Sommer-Larsen:2001, Wang:2007a}.  Since the
material is smooth below this scale, its evolution must be purely
sourced from larger scales, greatly simplifying the possible evolution. This suppression of
fluctuations below a finite wave number, that is captured by the dynamic range available
to the simulation, is thus the only difference with respect to $N$-body simulations of CDM structure 
formation.

A finite suppression scale should also lead to a reduced abundance of small haloes. These simulations
are of particular interest here, because by construction there is an easily resolvable minimum mass scale in the problem. 
Given this minimum scale, one might hope to resolve all the relevant scales and quickly arrive at converged
numerical solutions with increasing resolution.  However, it has been shown in  great detail that this is not 
the case~\citep[e.g.][and references therein]{Wang:2007a}. Contrary to predictions, it is observed
that nevertheless small haloes form and that they eventually merge to produce bottom-up hierarchical
structure formation on scales where this is not to be expected. Unphysical artificial fragmentation
of filaments into haloes that align like beads on a string are always present at scales of the initial mean 
particle separation. This finding is largely unaffected by different choices for the initial particle distribution
used to initialize the simulations. Uniform cubical lattice, quaquaversal tiling\citep{Conway:1998, Hansen:2007} 
or glass~\citep{White:1994} initial grids all produce these artefacts. While there was some
confusion initially whether these are real~\citep{Bode:2001} or whether they are absent in calculations that start from an initial
glass like distribution~\citep{Gotz:2003}, \cite{Wang:2007a} clearly showed these to be artificial both through direct hot/warm dark matter
simulations and idealized test calculations of sheets and filaments. 
It might seem tempting to circumvent this obvious numerical problem simply by an 
increase of resolution and by focusing on haloes above a more
conservative minimum number of particles than simulations of CDM. Such an approach
obviously shifts the fragmentation scale to smaller masses but clearly does not alleviate the 
associated problem that the fragments can merge to produce larger haloes. It  is thus 
unlikely that such an approach will quickly and reliably approach the correct solution.
Since simulation techniques that
avoid such spurious fragmentation are not at hand, \cite{Lovell:2011}, e.g., have filtered
their halo catalogues a posteriori. While such an approach avoids the obvious 
inclusion of numerical artefacts in the analysis, it does not guarantee that the 
fragmentation and associated relaxation effects have not impacted the dynamics
of the dark matter fluid in objects considered genuine.

While numerous numerical cosmology codes have been developed, they tend
to only differ in how the gravitational forces between particles are
computed. They all rely on the same assumption that
the problem to solve is a system of $N$ bodies attracting each other by their
mutual gravitational forces. Hence, potential errors introduced by this
assumption would be common to all these codes. 
The artificial fragmentation in warm/hot dark matter simulations is an
easily detectable and obvious error, demonstrating that N-body techniques
applied in the standard way can fail quite spectacularly and that convergence studies
improve the results only with a dramatic increase in the required computational
resources \citep{Wang:2007a}. One is thus prompted to wonder whether these the 
only important errors, or whether there are more that just escape detection. 
It is hence clearly desirable to develop
alternative techniques for cosmological collisionless simulations to
enable us to gauge the possible systematic biases and errors
introduced by the traditional $N$-body method.

In this paper, we introduce a numerical method that modifies the
traditional particle--mesh approach used in most modern cosmological
structure formation codes. Because of its lack of large dynamic range, the 
particle--mesh technique is now mostly used to compute only long-range interactions
in modern Tree-PM codes, but is very closely related to techniques used
in adaptive mesh codes.  The
particle--mesh method is one of the oldest and best understood
techniques in the field and was originally developed in the context
of plasma physics as described in great detail in the seminal
monograph by~\cite{Hockney:1981}. In their section on astrophysical
applications they impress upon the reader the importance of very large
particle numbers to achieve a collisionless situation.  In modern
plasma simulations, e.g., typically some hundred particles per mesh
cell are used \citep[e.g.][]{Chang:2008}. In cosmological
applications, however, this ratio is typical of order one to one
tenth~\citep[e.g.][]{Abel:2002c, Springel:2005a}. In hot plasmas, the
distribution function covers regions of the full six-dimensional phase
space. This is in contrast to the situation in CDM simulations where
the distribution function is at all times a three dimensional
submanifold of six-dimensional phase space. Only in regions of strong
mixing, this submanifold becomes manifestly six dimensional in a
space-filling sense. It is thus quite plausible that a smaller number
of particles might suffice to achieve a reasonable sampling of the
distribution function. However, the singular nature of gravity leads
to a very rapid growth of the volume of the dark matter sheet in phase
space. While some of this growth is automatically followed by the
Lagrangian motion of the $N$-body particles, their individual extent
is always assumed to be compact even though the solution shows that
they can be stretched to enormous volumes.  This is likely problematic
as it would require an enormous number of particles to 
faithfully sample the extraordinarily complex phase space structure.
The particle method only achieves this in a time-averaged sense.
We document the performance of the new technique using a set of well
defined test problems and hot/warm dark matter simulations and
contrast it to the standard particle mesh-algorithm. 

The structure of this paper is as follows: In Section \ref{sec:new-approach},
we present our new approach.
It is directly inspired by the tetrahedral decomposition of the
dark matter sheet in phase space that we presented in \cite{AHK11}. 
In Section \ref{sec:test-problems}, we apply our approach to various
simple test cases and compare the results with the literature, before 
in Section \ref{sec:HDM-simulations}, we apply the method to hot dark
matter simulations with a well-resolved cut-off in the initial
perturbation spectrum. In all cases, we compare the results with a traditional
$N$-body particle-mesh approach and quantify as well as discuss
the differences. Finally, we discuss our results in Section
\ref{sec:discussion} and conclude in Section \ref{sec:summary}.


\begin{figure}
\begin{center}
\includegraphics[width=0.48\textwidth]{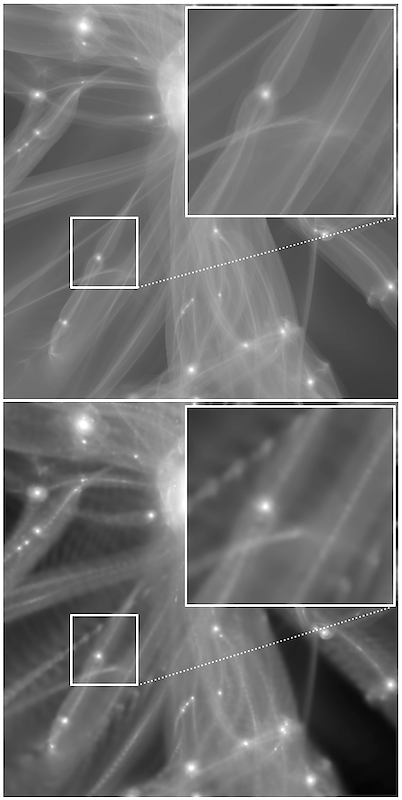}
\end{center}
\caption{Comparison of the density estimates from a WDM simulation obtained with the projection of tetrahedra (top panel; 
using the rendering method described in AHK12 and \citealt{Kaehler:2012})
as well as with an adaptive (SPH-like) smoothing approach (bottom panel). {\em Identical simulation particle data is used in both renderings.
The fragmented structure of the filaments disappears entirely if the tetrahedral elements are used.}}\label{fig:teaser}
\end{figure}


\section{A new approach to simulating cold self-gravitating collisionless fluids}
\label{sec:new-approach}

In this section, we describe a new approach to solving the equations of motion
of a cold collisionless dark matter fluid. Our approach is very closely related to the method
we proposed in \cite{AHK11} (AHK12 from here on) to trace the dark matter sheet in phase space
using a tetrahedral decomposition. The main idea of our approach is to use density estimates based
on this decomposition to compute gravitational forces.

\subsection{Motivation}
In \cite{Kaehler:2012}, we provided a detailed comparison of density field projections
obtained using the phase space sheet and conventional approaches (such as fixed size kernel
smoothing, adaptive kernel smoothing and Voronoi based estimates). In particular, we found 
there that the granularity of density projections completely vanishes when the density field is
rendered using the phase space sheet information. We show a similar comparison in
Figure~\ref{fig:teaser}, where the projected density is shown using the phase space sheet
(top panel) and using an adaptive SPH-kernel approach (bottom panel). Both images use the
same simulation data which was obtained from a cosmological $N$-body simulation with
truncated initial power spectrum and relatively low force resolution. The elimination of small-scale
perturbations prevents halo formation on small scales and leads to the emergence of prominent caustics.
In the adaptive SPH rendering,
we clearly see fragmented regions (see also the inset) that are not present when the tetrahedra 
are rendered. It is clear that if such a density field were used to compute forces, the visible
clumps would act as potential minima, attracting particles and leading to further fragmentation
of the filaments. Also, an inspection of the density field based on the tetrahedra clearly shows that 
fragmentation occurs in regions (filaments) that are strongly anisotropically compressed. Hence, an isotropic
kernel softening, even when adaptive, will not be able to capture the intricate and highly anisotropic
structure of caustics in these simulations. This has also been e.g. the motivation for \cite{Pelupessy:2003}
to discuss Delaunay triangulations instead of SPH kernel based estimates in
Lagrangian hydrodynamics. The convergence of an isotropically smoothed field to the one using the tetrahedra
is thus expected to be rather slow with increasing particle number.

In this paper, we will discuss a new particle-mesh method to construct a novel type of $N$-body solver
that separates flow tracers and mass tracers based on a decomposition of the phase space sheet into tetrahedral cells. 
The mass distribution in these cells is then approximated using pseudo-particles at monopole as well
as at quadrupole order, and the pseudo-particles are deposited using standard charge assignment 
to a fixed resolution mesh.
The density estimate obtained this way will serve as the source term for the gravitational forces.
We demonstrate that using the tetrahedral decomposition provides a density estimate that more accurately
reflects anisotropy on small scales and greatly reduces the growth of noise and artificial fragmentation
in $N$-body simulations.

\subsection{Outline of the method}
The aim is to solve the collisionless Boltzmann equation for a self-gravitating fluid
in an expanding universe, i.e. the 
evolution equation for the phase-space density $f(\mathbf{x},\mathbf{p},t)$
\begin{equation}
0=\frac{{\rm d} f(\mathbf{x},\mathbf{p},t)}{{\rm d}t} = \frac{\partial f}{\partial t} + \frac{\mathbf{p}}{ma^2}\cdot\boldsymbol{\nabla}_{x}f - m\boldsymbol{\nabla}_{x}\phi\cdot\boldsymbol{\nabla}_{p} f, 
\label{eq:boltzmann}
\end{equation}
supplemented with Poisson's equation for the gravitational potential $\phi$
\begin{equation}
\boldsymbol{\nabla}_{x}^{2} \phi = \frac{4\pi G m}{a}\left[ \int {\rm d}^3p \, \left( f -  \int {\rm d}^3x\, f \right) \right],
\label{eq:Poisson}
\end{equation}
where $m$ is the particle mass, $G$ is the gravitational constant and $a$ is the cosmological scale factor that
itself obeys the first Friedmann equation. We note that, while we perform the calculation in an
expanding universe, all that follows is perfectly valid for general cold collisionless fluids.

The full phase space distribution function $f(\mathbf{x},\mathbf{p},t)$
is manifestly six-dimensional and thus solving eq.~(\ref{eq:boltzmann}) as a partial differential equation for $f$
on a six-dimensional domain with reasonably high resolution
is computationally prohibitive, although it has been recently performed by \cite{Yoshikawa:2012} on a $64^6$ grid. 
For this reason, and since most of the phase space is filled sparsely in the case of cold fluids, most traditional 
methods resort to the $N$-body method and sample the distribution function $f$ at $N$ discrete points in space, i.e.
\begin{equation}
f(\mathbf{x},\mathbf{p},t) = \sum_{i=1}^N \delta_D\left(\mathbf{x}-\mathbf{x}_i(t)\right)\,\delta_D\left(\mathbf{p}-\mathbf{p}_i(t)\right),
\label{eq:pointwise_psdf}
\end{equation}
such that the Vlasov-Poisson equation
is fulfilled in terms of mass and momentum conservation equations for the particles. It can be hoped that the true
distribution function is then approximated by the particles in a time averaged sense. The particle discretization leads to a 
Hamiltonian $N$-body system whose phase space density $f$ is conserved along the characteristics $\mathbf{x}_i(t)$, 
$\mathbf{p}_i(t)$ defined by 
\begin{eqnarray}
\frac{{\rm d}\mathbf{x}_i}{{\rm d}t} &=& \frac{1}{m\,a^2}\mathbf{p}_i, \\
\frac{{\rm d}\mathbf{p}_i}{{\rm d}t} &=& -m\left.\boldsymbol{\nabla}_x \phi\right|_{\mathbf{x}_i}.
\label{eq:equations_of_motion}
\end{eqnarray}
We will also follow this approach but devise a novel technique to evaluate the force-term that takes into
account the continuous nature of the distribution function in phase-space
 from which the discrete number of characteristics is taken.

In classical  $N$-body methods, the source term in
Poisson's equation~(\ref{eq:Poisson}) is simply a sum over the particles that occupy discrete points
in phase space, e.g. $\mathcal{Q}_i\equiv(\mathbf{x}_i,\mathbf{p}_i)\in\mathbb{R}^6$ for particle $i$; i.e.
\begin{equation}
\boldsymbol{\nabla}_{x}^{2} \phi = \frac{4\pi G m}{a}\left[ \sum_{i=1\dots N} \delta_D(\mathbf{x}-\mathbf{x}_i) - \frac{N}{V}\right],
\end{equation} 
where $\delta_D(\cdot)$ is the Dirac $\delta$-function and $V$ is the simulation volume. Typically, for particle-mesh
methods, the $\delta_D$ are substituted for mass assignment functions that deposit the mass of a particle into one or
more grid cells, typically using either the Nearest-Grid-Point (NGP), Cloud-In-Cell (CIC) or Triangular-Shaped-Cloud (TSC) schemes
leading to increasingly higher smoothness of the gravitational forces (NGP, e.g., yields only a continuous force, not a differentiable one).
On the other hand, for mesh-free tree-based or direct summation based solvers, the $\delta_D$ are substituted for density
kernels of a finite extent that correspond to a softening of the gravitational force. Such a force softening is introduced somewhat ad-hoc
motivated by the fact that each particle does not actually represent a point-mass but rather a phase-space volume element
so that two-body collisional effects ought not be present.

Here, we propose a different technique. Based on the approach given in AHK12 to reconstruct the 
three-dimensional sheet of dark matter, we attempt to use a better estimate of the phase space distribution
function $f$ for use with standard $N$-body methods. For this reason, the source term in Poisson's equation should
build on a marginalisation of the phase-space density over the momentum coordinates. We will use a tetrahedral
decomposition of the dark matter sheet inspired by the approach from AHK12 for this purpose (see also \citealt{Shandarin:2011}). 
We describe the details of this approach in what follows.

\subsection{Decomposing the dark matter sheet}
Instead of stochastically sampling the phase space distribution function with particles that
do not carry any information about the continuous nature of the distribution function in standard $N$-body methods,
we explore here the piecewise linear approximation demonstrated in AHK12 to yield
a much improved density field that is defined everywhere in space.
For cold fluids, the phase space distribution function is an $n$-dimensional submanifold of
$2n$-dimensional phase space. A piecewise linear approximation can thus be achieved by a 
decomposition of the phase space distribution function into the simplices of $n$-dimensional
space, i.e. straight lines in 1D, triangles in 2D and tetrahedra in 3D. We will only consider
the three-dimensional case here and resort to a one-dimensional discussion only for
illustrative purposes.

The first step consists in setting up a volume decomposition for the three~dimensional
sheet on which a cold, i.e. without intrinsic velocity dispersion, dark matter fluid lives.
Initial conditions for our simulations are generated by displacing
particles from a uniform lattice according to the Zel'dovich approximation~\citep{ZelDovich:1970}.
The initial uniform lattice thus corresponds to the limit of very early times where (by definition)
no shell crossing has occurred and the velocity field is single valued. We can perform
a simple decomposition of this volume into tetrahedral cells that use the particle
positions on the lattice as their vertices. The advantage of using a uniform lattice
is that the connectivity information for the volume decomposition can be easily
recovered from the particle IDs which are set-up so that they simply encode the
Lagrangian coordinates of the particles. Using an initial glass-like distribution on the other
hand requires explicitly to generate the tetrahedra using e.g. a Delaunay triangulation
and to store the connectivity information. 

There are however many possible tetrahedral decompositions of the unit cell
of our uniform cubical lattice. We will consider two different decompositions in this paper. 
Note that the projected tetrahedra are always 
either convex or degenerate, with the degenerate state only occurring for an infinitesimally 
short time, when the tetrahedron is inverted due to shell-crossing and its volume changes sign. 
Also, mass is conserved by definition as the number of tetrahedra remains constant throughout
the simulation.

\subsubsection{The tesselating cubical decomposition}

The minimal decomposition of the unit cube
consists of five tetrahedra of two different sizes~\citep[cf.][]{Shandarin:2011}.
This tessellation is interesting since it is unique. It is however not isotropic and has two 
different masses associated with it, and for this latter property we do not consider it 
further here. Instead we use as one possibility the decomposition into six tetrahedra 
(equivalent to the Delaunay decomposition of the unit cube) that we give in AHK12 
and that has equal volume tetrahedra but is also anisotropic. 
It is shown by the shaded tetrahedra in the right panel of Figure~\ref{fig:particle_approx}.

\subsubsection{The non-tesselating octahedral decomposition}

\begin{figure}
\centerline{\includegraphics[width=0.4\textwidth]{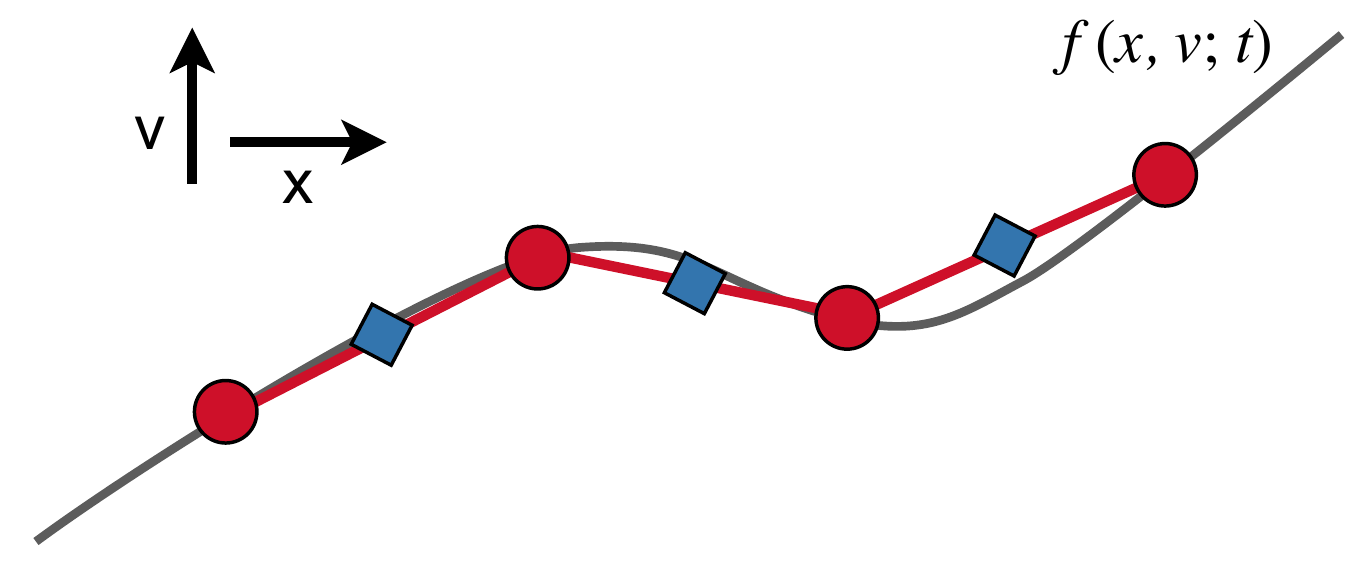}}
\caption{ The 1+1 dimensional case: The phase space distribution function $f(x,v;t)$ is sampled in standard $N$-body 
methods at discrete locations (red) but no information about the phase space structure (dark gray) is
retained. In the new approach, the connectivity is maintained (red lines), approximating
the true distribution function at linear order. In the TCM method discussed in this paper, we deposit the mass at zeroth
order at the centroid locations of the simplectic elements (blue diamonds). This leads to
a particle method with two types of particles: ``flow tracers'' (red) and ``mass tracers'' (blue).}\label{fig:linearapprox}
\centerline{\includegraphics[width=0.23\textwidth]{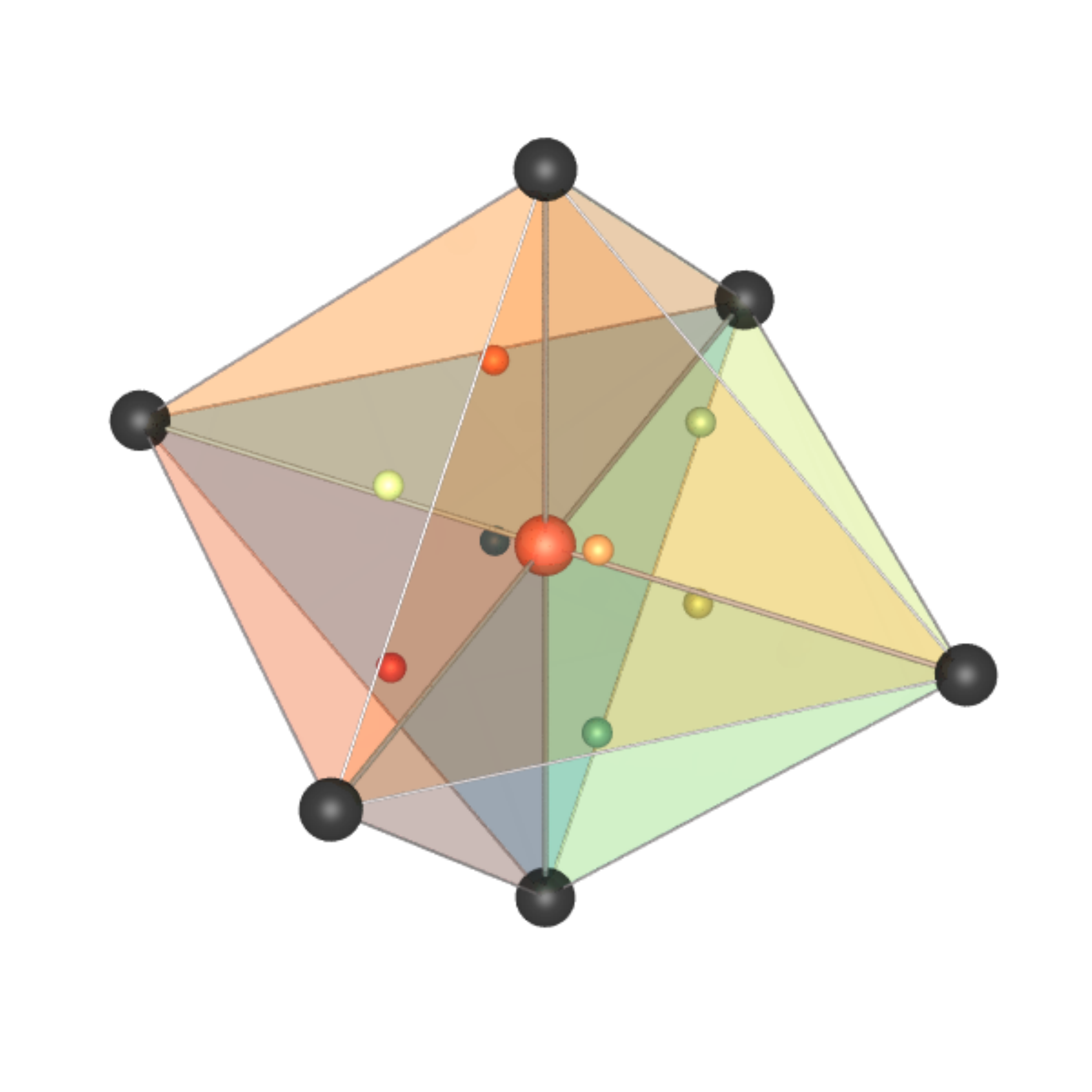}\includegraphics[width=0.26\textwidth]{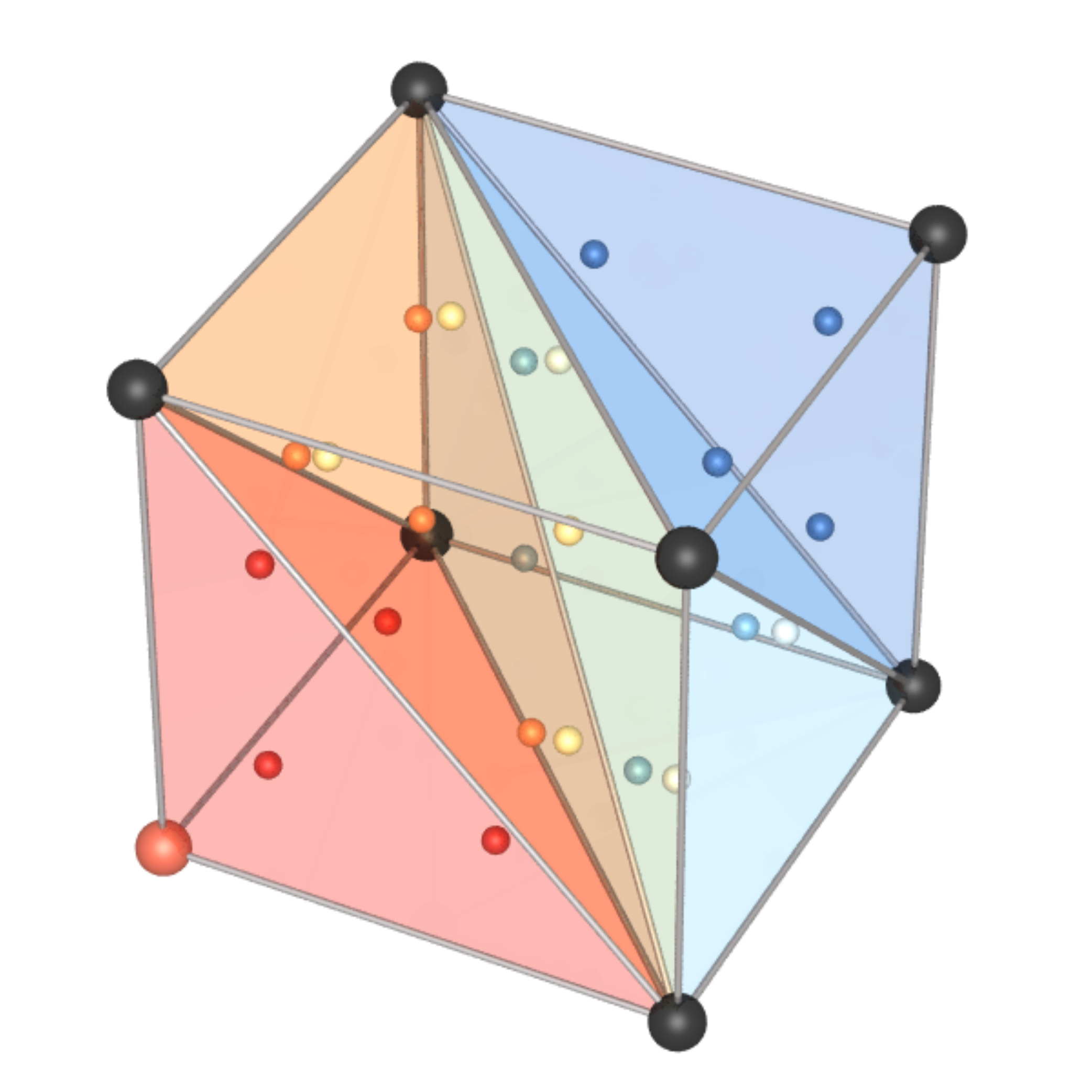}}
\caption{{\bf Left:} In TCM, every particle (large red) estimates its phase space volume from
  the octahedron it spans with its six nearest neighbours (large black) on the dark
  matter sheet. This shape is described by eight tetrahedra with
  centroids (indicated by the same colours) that give natural positions to think of the mass of
  the tetrahedron to be located in the monopole pseudo-particle approximation. 
  The centroids carry one eighth of the particle mass each. 
 {\bf Right:} In T4PM, we use the Delaunay triangulation of the unit cube associated with each particle (large red)
and its neighbours (large black) into six tetrahedra. The moment of inertia tensor of each tetrahedron can be
matched with 4 particles giving accuracy up to quadrupole order. The tetrahedra and their four-particles 
approximation are shaded in
the same colour. Each mass tracer particle thus carries $1/24$ of the particle mass.
}\label{fig:particle_approx}
\end{figure}

In addition, we also consider an alternative decomposition for which we give up the 
requirement that the tetrahedra provide also a tesselation of the cubical lattice. 
Instead they are to not induce any intrinsic anisotropies that are not already present 
in the cubical lattice. 
We thus construct a set of tetrahedral cells whose union is a octrahedron centred on
 each particle as illustrated in Figure~\ref{fig:particle_approx}. This does
 not provide a tesselation of space as the octahedra overlap and thus each 
 point in space belongs to two tetrahedra that themselves belong to two different
 octahedra. This configuration naturally gives a configuration in which 
 the centroids of the eight tetrahedra associated with each particle correspond
 to a cartesian mesh refinement with a refinement factor of two of the initial cubical
 particle lattice. 
 Since, in contrast
to the cubical decomposition, now each tetrahedron contributes
twice, we obtain a mass density that has to be simply divided by two to obtain the 
correct mass density in configuration space. Note that this has the additional advantage that
always information from several tetrahedra contributes to the density estimate. The 
tetrahedra from this decomposition are shown in the left panel of Figure~\ref{fig:particle_approx}
as shaded volumes.

\subsection{Multipole expansion and pseudo-particle approximation of the tetrahedral mass elements}
\label{sec:multipole}

 We want to use the density field $\rho(\mathbf{x})=m\int{\rm d}^3 v\,f$, construed by projecting the tetrahedra into
 configuration space, as the source term for Poisson's equation~(\ref{eq:Poisson}). As we have argued in AHK12, 
 the connectivity of the volume decomposition does not
change once the dark matter sheet is evolved since the sheet never intersects itself in
six dimensional phase space nor can discontinuities arise, i.e. the sheet cannot tear.
 In order to solve Poisson's equation for the gravitational potential and to obtain gravitational forces, 
 one possible approach is to use the analytic potential produced by each tetrahedron \citep[e.g.][]{Waldvogel:1979} and to
 devise a Green's function approach, where the Green's function depends on the shape and orientation
 of each tetrahedron. This is likely computationally extremely expensive so that we will not
 pursue this approach here further.
Hence, rather than modelling the full mass distribution of each tetrahedron, we approximate it
 at monopole or at quadrupole order using pseudo-particles. This is less accurate, but computationally efficient
 and it will allow us to use established
  particle-based gravitational $N$-body Poisson solvers, such as the particle-mesh (PM) method \citep[e.g.][]{Efstathiou:1985}, 
  the tree method \citep[e.g.][]{Barnes:1986}, as well as the combination to tree-PM \citep[e.g.][]{Xu:1995}, 
  to compute force fields and potentials.  

Assume a homogeneous (i.e. uniform density $\rho={\rm const.}$) tetrahedron
$\mathcal{T}$, projected into configuration space and described by its four vertices
$(\mathbf{x}^{(1)},\mathbf{x}^{(2)},\mathbf{x}^{(3)},\mathbf{x}^{(4)})$ with $\mathbf{x}^i\in\mathbb{R}^3$.
Expanding it to monopole order is equivalent to concentrating all mass at
the centroid position $\mathbf{x}_c$ that is equivalent to the center of mass defined by
\begin{equation}
\mathbf{x}_c \equiv \frac{1}{M}\int_\mathcal{T} \rho \mathbf{x}\,{\rm d}^3 x = \frac{1}{4}\sum\mathbf{x}^{(i)},
\end{equation}
using that $M\equiv\int_\mathcal{T}\rho\,{\rm d}^3x$. See the left panel of Figure~\ref{fig:particle_approx} for the location
  of the centroids in our decomposition, and Figure~\ref{fig:linearapprox} for the 1-dimensional
  case where we can show all of phase space. {\em We abbreviate the combination of the octahedral decomposition
  with the approximation of the mass distribution up to monopole order as ``TCM'' in this paper.}
  
Higher order multipole moments of arbitrary mass distributions can be approximated using more particles 
\citep[see also e.g.][who however follow a different approach]{Makino1999}.  
In this paper, we want to approximate the mass distribution up to quadrupole order which requires four pseudo-particles.
The quadrupole moment is given by the moment of inertia tensor
\begin{equation}
{\rm Q}_{ij} \equiv \int_\mathcal{T} \rho \left( x_k x_k \delta_{ij}-x_i x_j\right)\,{\rm d}^3 x,
\end{equation}
which can be integrated over the tetrahedral domain to yield, e.g.,
\begin{eqnarray}
{\rm Q}_{xx} & = & \frac{M}{10} \sum_{i\leq j} \left[ y^{(i)}y^{(j)}+z^{(i)}z^{(j)} \right]\\
{\rm Q}_{xy} & = &-\frac{M}{20} \sum_{i,j} \left(1+\delta_{ij}\right)\, x^{(i)}y^{(j)} \\
&\dots&
\end{eqnarray}
We now want to place four particles of mass $M/4$ that match both the monopole and quadrupole
moment of the tetrahedron. We make the Ansatz that these particles reside at the locations
\begin{equation}
\mathbf{a}^{(i)}\equiv \alpha \mathbf{x}^{(i)} + \beta \mathbf{x}_c.
\end{equation}
Matching the monopole requires that $\alpha+\beta=1$, implying that $\mathbf{a}$ must lie
somewhere on the line connecting the vertex and the centroid. Calculating the moment of
inertia tensor for the system of four particles and matching it to the inertia tensor of
the tetrahedron yields 
\begin{equation}
\alpha = \frac{\sqrt{5}}{5}\simeq0.447\quad\textrm{and}\quad\beta = 1-\frac{\sqrt{5}}{5}\simeq0.553.
\end{equation}
By placing pseudo-particles at the four locations $\mathbf{a}^{(i)}$ it is thus possible to match both
the monopole and the quadrupole moment of the mass distribution of the homogeneous tetrahedron.
{\em We abbreviate the combination of the cubical decomposition with the quadrupolar pseudo-particle
approximation as ``T4PM''  in this paper}.

We will use both the monopole and quadrupole approximation in the remainder of the paper in order
to be able to constrain the influence of the approximation to the true tetrahedral mass distribution on
our results. In particular, we note that the centroid approximation to the mass distribution of a single
tetrahedron
is invariant under all linear deformations of the tetrahedron (rotation, shearing) with respect to the
centroid. These linear deformations are however correctly represented in the quadrupole approximation.

 \subsection{Mass deposition and Poisson solvers}
    
The main conceptual idea behind our tetrahedral-particle-mesh technique differs from previous 
cosmology codes used to study dark matter. This difference rests in that we separate mass tracers
from flow tracers. In standard $N$-body methods, the particles carry the mass, are thus the building
blocks of the density field which determines the gravitational potential and forces, while the particles
also correspond to Lagrangian fluid elements. Instead, here we use the $N$-body particles only
as flow tracers, while they generate tetrahedra that provide a volume based decomposition of the
full three-dimensional phase-space distribution function which is accurate at linear order. We then
approximate the mass distribution of the tetrahedra up to monopole order in TCM and up to 
quadrupole order in T4PM using pseudo-particles that serve as mass tracers. The mass of
each tetrahedron is thus spread out over eight ``mass tracer'' particles in TCM and 24 ``mass tracer''
particles in T4PM whose positions depend on
the location of the neighbouring particles on the dark matter sheet. These mass tracers are thus not
explicitly evolved. Since the connectivity information is conserved over time, they can be computed
from the flow tracer particles at any time.  

To compute the force generated by the ``mass tracers'' and experienced by the ``flow tracers'', we use a modified
version of the publicly available {\sc Gadget-2} code. {\sc Gadget-2} uses the tree-PM method
so that our modifications simply had to introduce a new particle type of ``mass tracers'' while the
standard particles are retained and assigned a zero mass.  The positions of the ``mass tracers'' are updated before
each force calculation. This update depends only on the positions of the flow tracers and the connectivity of 
the dark matter sheet (that however remains constant
over time) and can thus be simply encoded in their particle IDs. Since the ``mass tracers'' are uniquely determined
by the ``flow tracer'' positions, they could also be computed on the fly, so that no additional memory is needed.

\begin{figure}
\begin{center}
\includegraphics[width=0.48\textwidth]{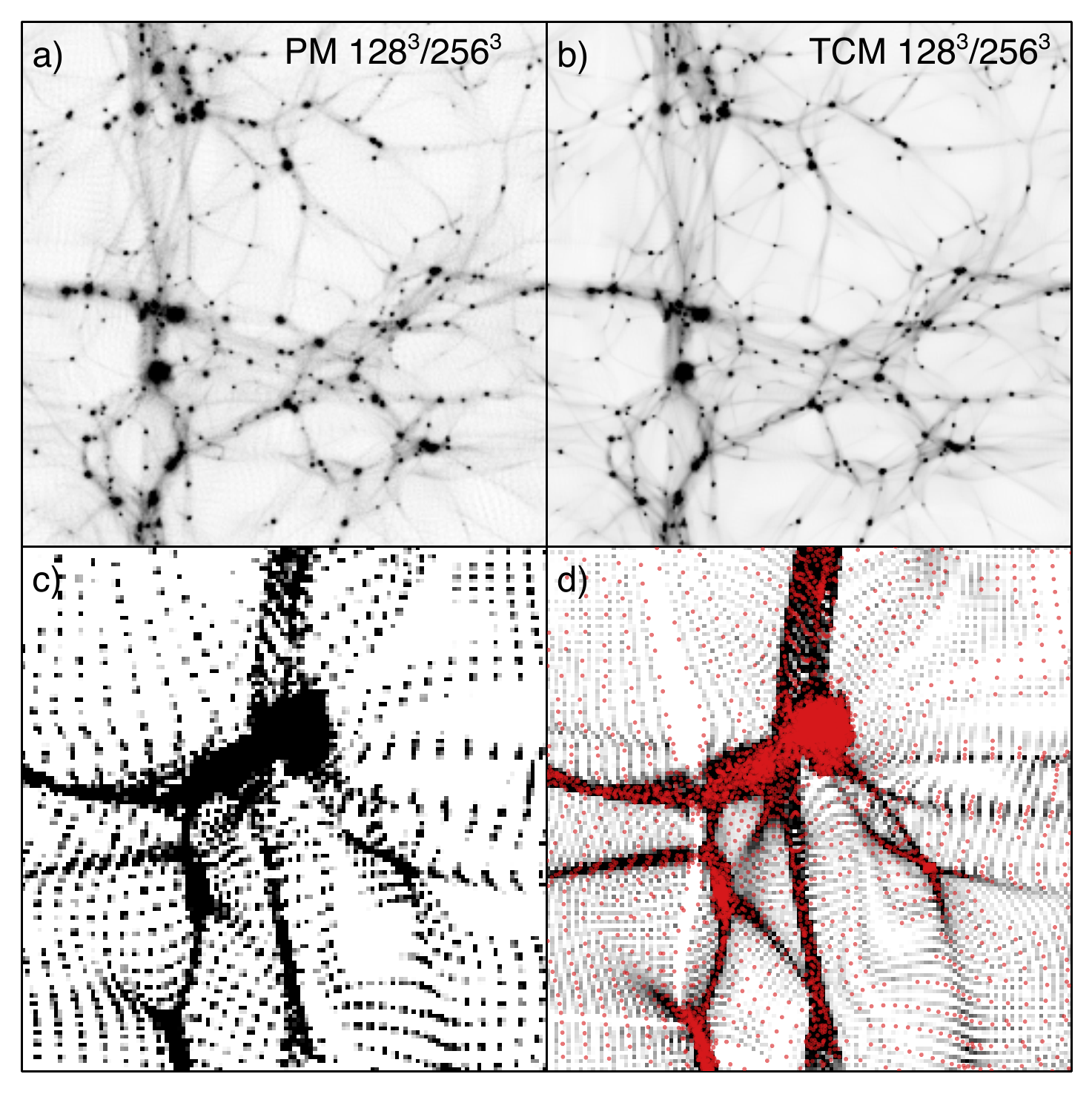}
\end{center}
\caption{Projections (top panels a and b) and slices (bottom panels c and d) through a
  $256^3$ particle mesh grid of $128^3$ particles in a warm dark
  matter simulation. Left panels (a and c) show the standard cloud-in-cell
  deposit and the right panels (b and d) again using CIC but at the eight times
  more numerous tetrahedra centroids which approximate the masses of the
  tetrahedra at lowest order.  The grey scale is logarithmic
  and identical for the slices and projections respectively. The
  slices show only a quarter of the full area of the simulation. In the
  bottom right panel we overplot the locations of the particle
  positions in red that contribute to both the CIC and TCM
  deposit. The tetrahedra centroids provide effective interpolation of
  the density field and their CIC density appears significantly less
  noisy with more contrast. The regular pattern originating from the
  original uniform lattice remains, but appears at higher spatial
  frequency in TCM.  }\label{fig:cic-slices}
\end{figure}

Standard CIC interpolation
is then used to deposit the mass tracer pseudo-particles at the centroid positions for the monopole approximation used in TCM
or the four positions needed for the quadrupole approximation in T4PM. This
deposits the mass of each tetrahedron into cubical cells of the size
of the underlying grid and obtains an improved density estimate over
the standard particle-mesh techniques.  Figure~\ref{fig:cic-slices}
shows slices and projections of the interpolated density arrived at
with this procedure in the TCM case and contrasts it to standard CIC for the same particle
distribution. Note that this is still a rather poor description of the
true underlying density field which would require to fully map the
tetrahedra onto the grid as shown in Figure~\ref{fig:teaser}. The latter is much more computationally
cumbersome than the simple pseudo-particle approximations we present here, where the mass tracers
are treated as particles whose mass get deposited
in the standard CIC fashion. The simple effective mass resolution
increase gives benefits much greater than a standard calculation with
8 or 24 times more particles as we will demonstrate next after we have
introduced the simulation parameters of our resolution study in the
following section. Note that the tree-part of the gravitational force (and
the CIC deconvolution) has
been switched off in our modified version of {\sc Gadget-2} so that we
are using it simply as a parallel implementation of the standard PM method, apart from
our modifications to the mass deposition based on the dark matter sheet.

\subsection{The piecewise linear approximation, its evolution and growth of errors}
\label{sec:linearerrors}
Apart from the approximations to the mass distribution of the homogeneous tetrahedra
discussed in Section~\ref{sec:multipole}, the other approximation that is used in
our new method is the piecewise linear approximation of the phase space distribution 
function $f(\mathbf{x},\mathbf{p},t)$ by the tetrahedral elements. We now 
discuss how the true distribution function relates to this approximation and 
how errors can arise.

The continuous phase space distribution function can be written analogously to eq.~(\ref{eq:pointwise_psdf})
as
\begin{equation}
f(\mathbf{x},\mathbf{p},t) = \int \delta_D(\mathbf{x}-\mathbf{x}_\mathbf{q}(t))\,\delta_D(\mathbf{p}-\mathbf{p}_\mathbf{q}(t))\,{\rm d}^3q,
\end{equation}
where $\mathbf{q}\in\mathbb{Q}=\left[0,1\right[^3$ (due to the periodic boundary conditions, in fact, $\mathbb{Q}$ is
the unit 3-torus) is a parameterisation of the hypersurface of the phase space sheet,
and can be thought of as a continuous generalization of the particle indices, i.e. a bijective mapping
$\mathbb{Q}\rightarrow\mathbb{R}^6\,:\,\mathbf{q}\mapsto(\mathbf{x}_\mathbf{q}, \mathbf{p}_\mathbf{q})$ exists at all times. 
Since $\mathbb{Q}$ is three-dimensional, the local tangent space spans only a three-dimensional subspace of six-dimensional phase space
and is given by the Jacobian $(\partial x_i/\partial q_j, \partial p_i/\partial q_j)$.

{\em The simplectic decomposition of the phase space sheet discussed in AHK12 provides a simplicial
approximation to the continuous phase space distribution function} $\mathbf{q}\mapsto f(\mathbf{q},t)$
\citep[e.g.][]{Alexandrov:1961}. Such a simplicial approximation to a continuous function is equivalent
to a piecewise linear interpolation across each simplex and allows us to compute
 finite difference estimates of the two tensors ${\partial x_i}/{\partial q_j}$ 
 and ${\partial v_i}/{\partial q_j}$, which are constant across each simplex, by computing the differences of vertex positions 
 and velocities along the edges of each simplex. As the $q_j$ are just the Lagrangian 
 coordinates in our case, $\partial q_j$ corresponds to neighbouring particles 
 in the initial conditions that are connected by an edge of a simplex.

The two tensors above can be combined to yield the rate-of-strain tensor
$\partial v_i / \partial x_j$. Note that this tensor is singular whenever $\partial x_i/\partial q_j$
has one or more zero eigenvalues which corresponds to shell-crossing along one
or more dimensions (see e.g. Figure~\ref{fig:lagrangian_velocity}, below, where
we show $x(q)$ and $v(q)$ at late stages of a plane wave collapse where multiple
caustics have arisen).

The evolution equations are analogously to the N-body evolution equations given by
\begin{equation}
\frac{{\rm d}\mathbf{x}_\mathbf{q}}{{\rm d}t} = \frac{\mathbf{p}_\mathbf{q}}{m a^2}\quad\textrm{and}\quad
\frac{{\rm d}\mathbf{p}_\mathbf{q}}{{\rm d}t} = -m \boldsymbol{\nabla}_x \phi.
\end{equation}

The decomposition of the phase space sheet into tetrahedra can be thought of as considering only a discrete set
of points $\mathbf{q}_i$ and approximating $\mathbf{x}(\mathbf{q})$ and $\mathbf{p}(\mathbf{q})$ linearly in
between, i.e. across the tetrahedral elements.  
Over time, differences between the piecewise linear approximation and the true distribution function can arise from 
second order terms in the force field and the fluid motion. Deviations from linearity are small if the flow field across 
a simplex is dominated by linear motion, i.e.  translation, shear and solid body rotation. The error of
the linear approximation is, at leading order, given by the second derivatives of the phase space 
sheet (related to the local curvature) described by the two tensors $\partial^2 x_i / \partial q_j\partial q_k$
and $\partial^2 p_i / \partial q_j\partial q_k$ and has the time derivative
\begin{eqnarray}
\frac{\partial}{\partial t}\frac{\partial^2 x_i}{ \partial q_j\partial q_k} & = & \frac{\partial^2}{\partial q_j \partial q_k}\frac{p_i}{m a^2}\\
\frac{\partial}{\partial t}\frac{\partial^2 p_i}{ \partial q_j\partial q_k} & = & -m \frac{\partial^2}{\partial q_j \partial q_k} \frac{\partial \phi}{\partial x_i}.
\end{eqnarray}
The right hand side of these two expressions describes the deviation from spatial linearity of the 
velocity field and the gravitational force across the simplex respectively.

\begin{figure}
\begin{center}
\includegraphics[width=0.47\textwidth]{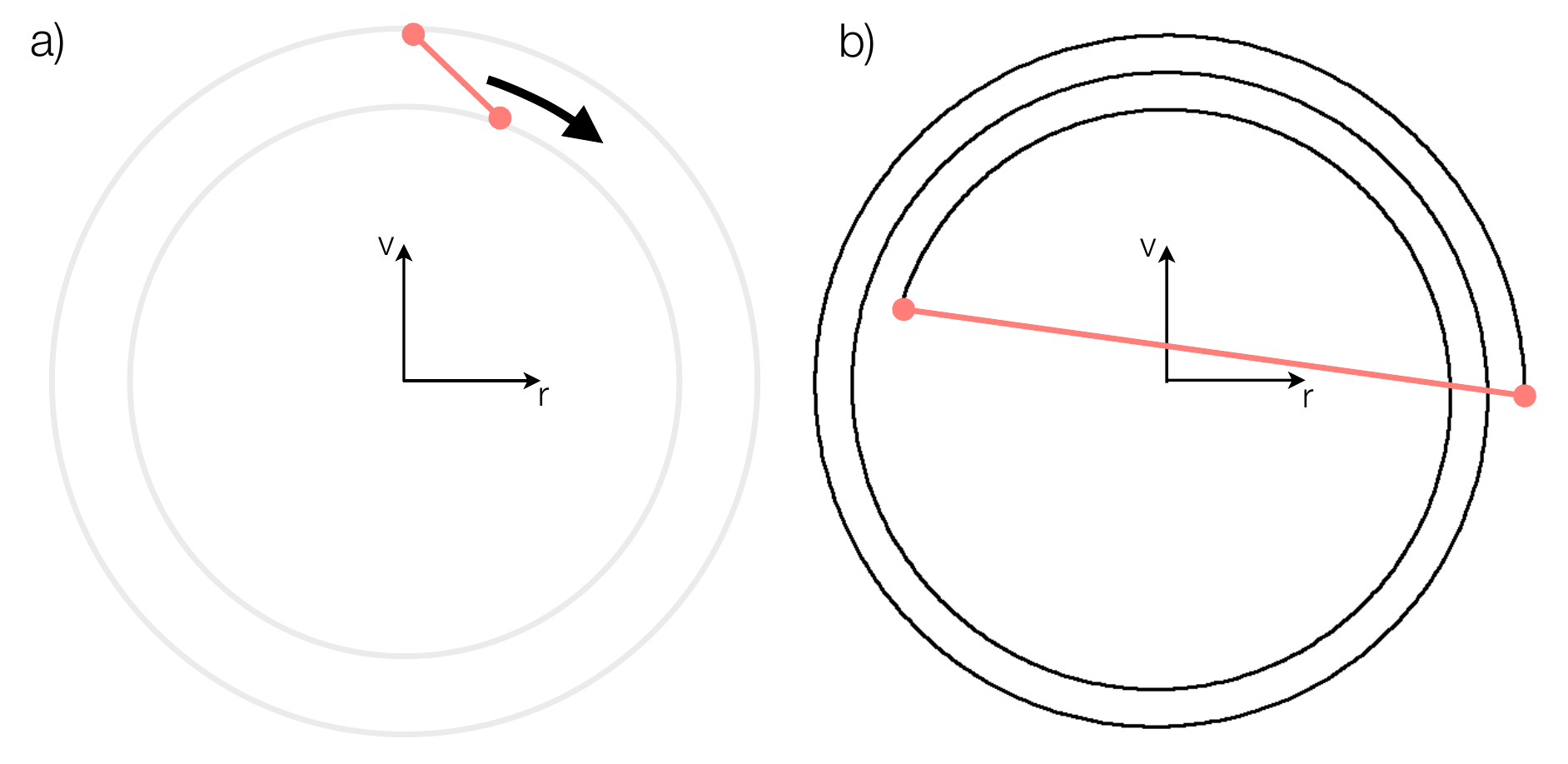}
\end{center}
\caption{\label{fig:mixing}The limitations of a piecewise linear approximation without refinement in regions of phase mixing. Consider
the linear element of the distribution function (red line element in panel a), orbiting in a central potential. After a few orbits, the line element will
cover the region between the two grey circles with a spiral (black line in panel b). Without refinement, the mass will be assigned to the stretched red
line element in b) rather than to the spiral.}
\end{figure}

The proposed method is thus expected to perform well whenever the velocity shear and gravitational tides are
constant across an element. If the piecewise linear approximation is exact initially, all errors will
arise from the change of the tidal field across the element and then propagate to the velocity shear. 
This error is negligible if the simplices are small, but errors will grow if the piecewise linear approximation 
cannot resolve changes in the 
tidal field, i.e. if the simplex becomes larger than relevant scales in the tidal field. At this point, subdivision 
of the simplex is necessary to control the errors. 

An example where the simplicial approximation breaks down quickly are regions of phase mixing. 
This situation is demonstrated in Figure~\ref{fig:mixing}.
A simplectic element orbiting in a fixed potential is stretched into a spiral. If no refinement is performed,
the simplex cannot capture the growth of the associated volume, leading to mass deposition that
is biased towards the centre of the potential.

A refinement criterion emerges naturally
by requiring an upper bound on the spatial derivatives of the velocity shear and tidal tensors. One way to
implement such a refinement criterion would be to locally compute the geodesic deviation equation
discussed by \cite{Vogelsberger:2011}. However, it can be achieved more
easily by computing velocities and accelerations also at the centroid positions and verifying
their deviation from the finite difference estimate obtained from the vertices. 

In this paper, we will not
discuss refinement further as it shall suffice here to introduce the method and investigate
its performance outside of those regions where loss of resolution is problematic. It is important however
to keep this limitation of our method in mind when interpreting some of the results in the remainder
of the paper. It will become clear from the results shown below that the 
Lagrangian motion of the particles is insufficient to track the full evolution in six dimensional 
phase space, as the dark matter sheet can grow exponentially fast during multiaxial gravitational
collapse leading to the densest regions in multidimensional simulations. In order to arrive at a method that 
works also in such regions dominated by strong mixing, additional
refinement is imperative. A detailed discussion and validation of such a refinement approach is
however beyond the scope of this first paper and will be deferred to a second 
paper, which is already in preparation.

\subsection{Computational Performance}
The computational cost for the new method is, as expected, larger for the same 
number of tracer and standard $N$-body particles, larger as expected, but no additional computational
complexity is introduced. For the runs shown in Sections \ref{sec:test-problems} and
\ref{sec:HDM-simulations}, we observe that T4PM requires about three times more
computing time than the standard PM runs. This overhead is caused by the on-the-fly
creation and communication of the mass carrying particles, when depositing the 
mass distribution onto the PM grid. We note that this overhead is linear in the number
of particles, i.e. $\mathcal{O}(N)$ and will thus be dominated by the cost of the force calculation for
large particle numbers which scales as $\mathcal{O}(N_g \log N_g)$ for an FFT of
$N_g$ mesh cells. Peak memory usage is for both TCM and T4PM also about 3 times
larger than for PM, due to additional pointers that we store for each tracer particle 
to be able to quickly create the mass carrying particles on the fly. We note that 
for T4PM we have effectively 24 times more particles to sample the density
field, so that an overhead of three appears small. We also note that a more optimal
solution is possible, where the connectivity is created on the fly with no
memory overhead over the PM method at the expense of very little extra CPU time. 
For these first tests, we did not perform this optimization.


\section{Low-dimensional test problems}
\label{sec:test-problems}
In this section, we investigate the validity and compare the performance of our new TCM and T4PM methods
 with that of the standard PM method using three test problems of increasing complexity. In particular, we want to study the emergence
of spurious two-body effects that signal deviation from the collisionless limit with increasing force resolution. We want to stress
that the new method applies to modelling cold collisionless systems in the fluid approximation, and is thus rather different in the
character of test problems that apply, compared to the standard $N$-body method that is also suited to study hot systems 
such as star clusters or few body problems. The test problems in this section are thus all collapse problems arising from 
one and two~dimensional perturbations of the dark matter sheet.
All simulations in this section assume an Einstein-de~Sitter cosmology ($\Omega_m=1$, $\Omega_\Lambda=0$), Hubble parameter $h=0.7$, 
starting redshift $z_{\rm start}=100$ and box length of $L=10\,h^{-1}{\rm Mpc}$.

We note that, while all test cases investigated in this section are one- and two-dimensional, we perform
the simulations in three dimensions. In all cases, initial conditions are given by an initial gravitational 
potential which varies in only one or two dimensions. 
The initial particle coordinates and velocities are then determined from this potential using the Zel'dovich 
approximation. This converts the gravitational potential into a velocity potential leading to an irrotational
flow. The tesselation of the dark matter sheet is performed on the unperturbed cubical lattice (as in AHK12).

\subsection{Axis-parallel plane wave collapse}
\label{sec:test_axis_plane_wave}
We first consider the standard plane wave collapse, i.e. 
a one-dimensional sinusoidal potential perturbation
\begin{equation}
\phi(\mathbf{x}) = \bar{\phi}\cos(k_p x),
\end{equation}
where $\mathbf{x}=(x,y,z)$, $k_p = 2\pi/L$ and $\bar{\phi}$ is chosen so
that shell crossing occurs at an expansion factor of $a_c=1/7.7$. 
Initial particle positions and velocities are obtained
by applying the Zel'dovich approximation \citep{ZelDovich:1970} to
an unperturbed regular Cartesian lattice of particles. The plane wave thus obeys
the symmetry of both the initial particle distribution and of the mesh structure used to
compute the forces. At no time accelerations in the y- or z-direction can arise.

For this one-dimensional problem, the Zel'dovich approximation provides the exact
solution until shell-crossing occurs \citep[e.g.][]{Shandarin:1989},
allowing us to study the numerical errors in the mildly non-linear stage.

\begin{figure}
\centerline{\includegraphics[width=0.47\textwidth]{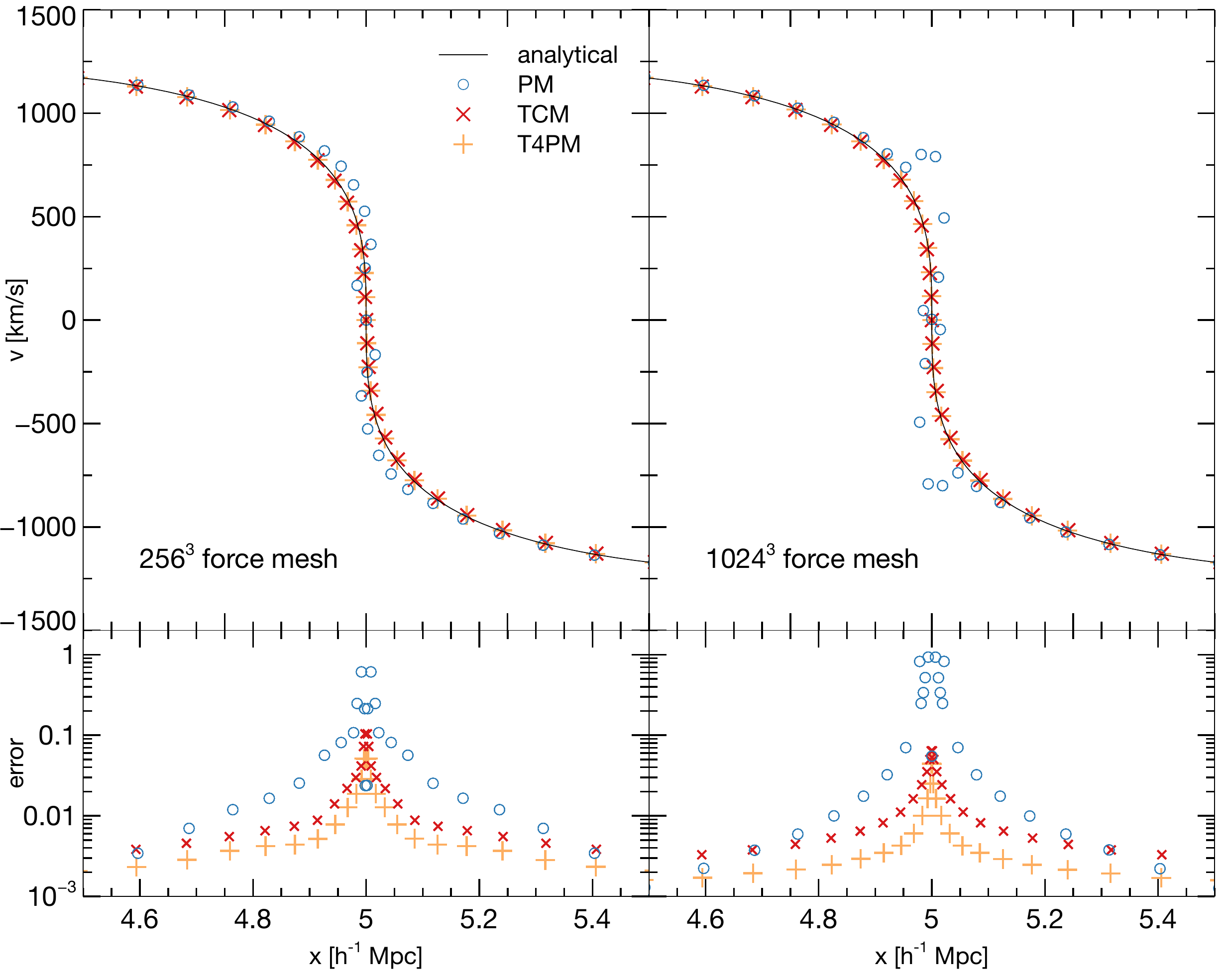}}
\caption{\label{fig:parallel_wave_crossing}Central region of the {\em axis-parallel} plane wave at the time of shell crossing. The upper panels show the phase space structure. Shown are
the results for the standard PM method with $64^3$ particles (blue circles) and for the TCM  (red crosses) and T4PM (yellow crosses) methods with $64^3$ tracers.
The analytic solution is indicated by the thin black line.
The left panels show results for a $256^3$ mesh to compute forces, the right panels for $1024^3$ cells. The lower panels show the 
absolute value of the velocity error in units of the RMS velocity. The impact of collisionality and its dependence on force resolution
for the standard particle method is clearly visible already at this early time.
}
\end{figure}
\begin{figure}
\centerline{\includegraphics[width=0.4\textwidth]{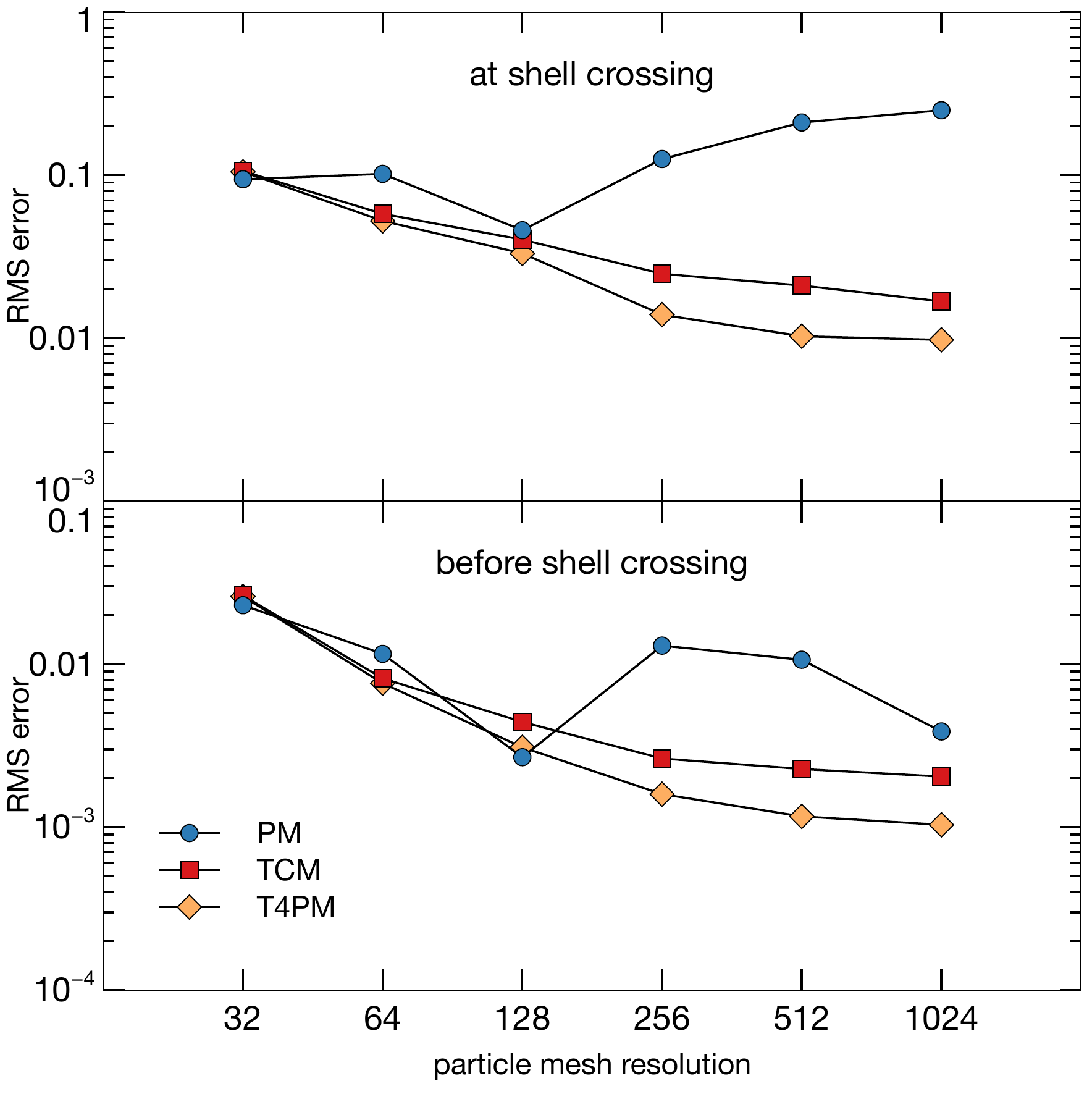}}
\caption{\label{fig:wave_convergence}Error of the numerical solutions with respect to the analytical
solution before and at shell crossing in the {\em axis-parallel} plane wave collapse problem. 
We show the behaviour of the RMS errors in the velocity in units of the RMS velocity 
with changing force resolution. Results are
given for standard PM with $64^3$ particles (blue circles) and for TCM (red squares) and T4PM (yellow diamonds) with $64^3$ tracers. 
The mesh used to compute the forces is varied from
$32^3$ to $1024^3$ cells. The errors are given at two specific times of the collapse: at shell-crossing, as well as at an
expansion factor of $1.3$ before the shell-crossing.
}
\end{figure}

In Figure~\ref{fig:parallel_wave_crossing}, we show the two-dimensional phase-space structure
of the central region of the wave at shell crossing. In all cases $64^3$ particles were used (i.e. $64^3$ flow
tracers for TCM and T4PM). The 
right panel shows the results for the highest force-resolution that we employed, 
while the left panel shows the corresponding results for
a lower force resolution. We observe that when high force resolution is used, two-body 
effects in standard PM lead to significant spurious accelerations of the particles. Note that
the $1024^3$ mesh corresponds to a mere $1/16$ of the mean particle separation, still significantly below
the $1/30$--$1/60$ typically used in cosmological simulations. For both TCM and T4PM we see no evidence
for such two-body effects at the force resolutions investigated here, the error with respect to
the analytic solution is about an order of magnitude lower in the innermost region and generally
about a factor of 2-3 lower for T4PM than for TCM. We note however that
the error with respect to the analytic solution does not drop as fast as in the standard PM case at 
larger distances but approaches a roughly constant value of $\sim 1-2\times10^{-3}$. This reflects the
fact that the tetrahedra sample volumes rather than a collection of point-like particles so that 
accelerations are expected to deviate somewhat when local particle separations become larger than the 
mean particle separation.

We quantify the deviation from the analytic solution as a function of force resolution in Figure~\ref{fig:wave_convergence}.
The RMS errors of the numeric solution with respect to the analytic solution are shown in units
of the RMS velocity at two times: before shell crossing
at $a/a_c=0.77$ and at shell crossing $a=a_c$. We notice a remarkable and significant difference between 
our new method and standard PM. For the standard PM method, the errors are smallest when the force resolution is about half the
mean particle separation. If the force resolution is increased, the errors become significantly larger. The result
for TCM and T4PM is completely different. Here we observe a monotonous reduction of the error with increasing force
resolution. Before shell-crossing the minimal errors achieved by the two methods are comparable
at $\sim 4\times10^{-6}$, but the analytic solution is smooth so that no high force resolution is necessary to
resolve it. At shell crossing, higher force resolution is necessary due to the presence of a sharp feature in the 
solution to achieve small errors. We thus observe that the smallest error is achieved with T4PM and TCM at the
highest force resolution. This shows the dilemma of standard PM: high force resolution is needed to capture
small features in the physical solutions, but at the same time high force resolution increases two body effects.
At the force resolutions investigated we do not observe that either TCM or T4PM suffers from comparable problems.

\begin{figure*}
\centerline{\includegraphics[width=0.8\textwidth]{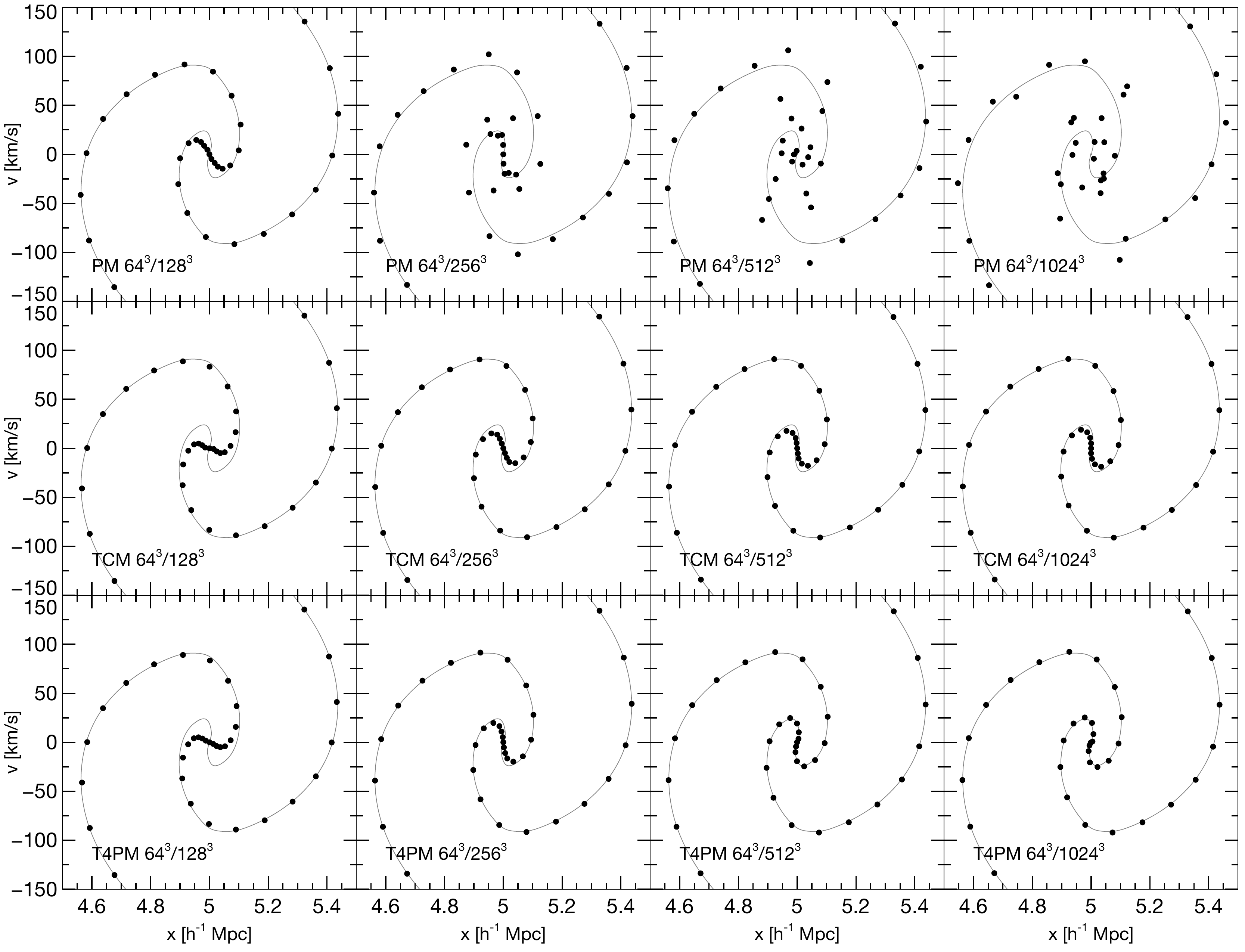}}
\caption{\label{fig:plane_parallel_late}Central region of the {\em axis-parallel} plane wave at an expansion factor of 7.7 after shell crossing. Shown are
the results for the standard PM method with $64^3$ particles (top row) and for the TCM (middle row) and T4PM (bottom row) method with $64^3$ flow tracers .
From left to right, the columns show results for an increasing force resolution from $128^3$ to $256^3$ to $512^3$ to $1024^3$ mesh cells.
 The thin grey line is the solution obtained with standard PM with $512^3$ particles and
$512^3$ mesh cells. Two body collisions destroy the spiral in the standard PM case for higher force resolutions.}
\end{figure*}

\begin{figure}
\centerline{\includegraphics[width=0.47\textwidth]{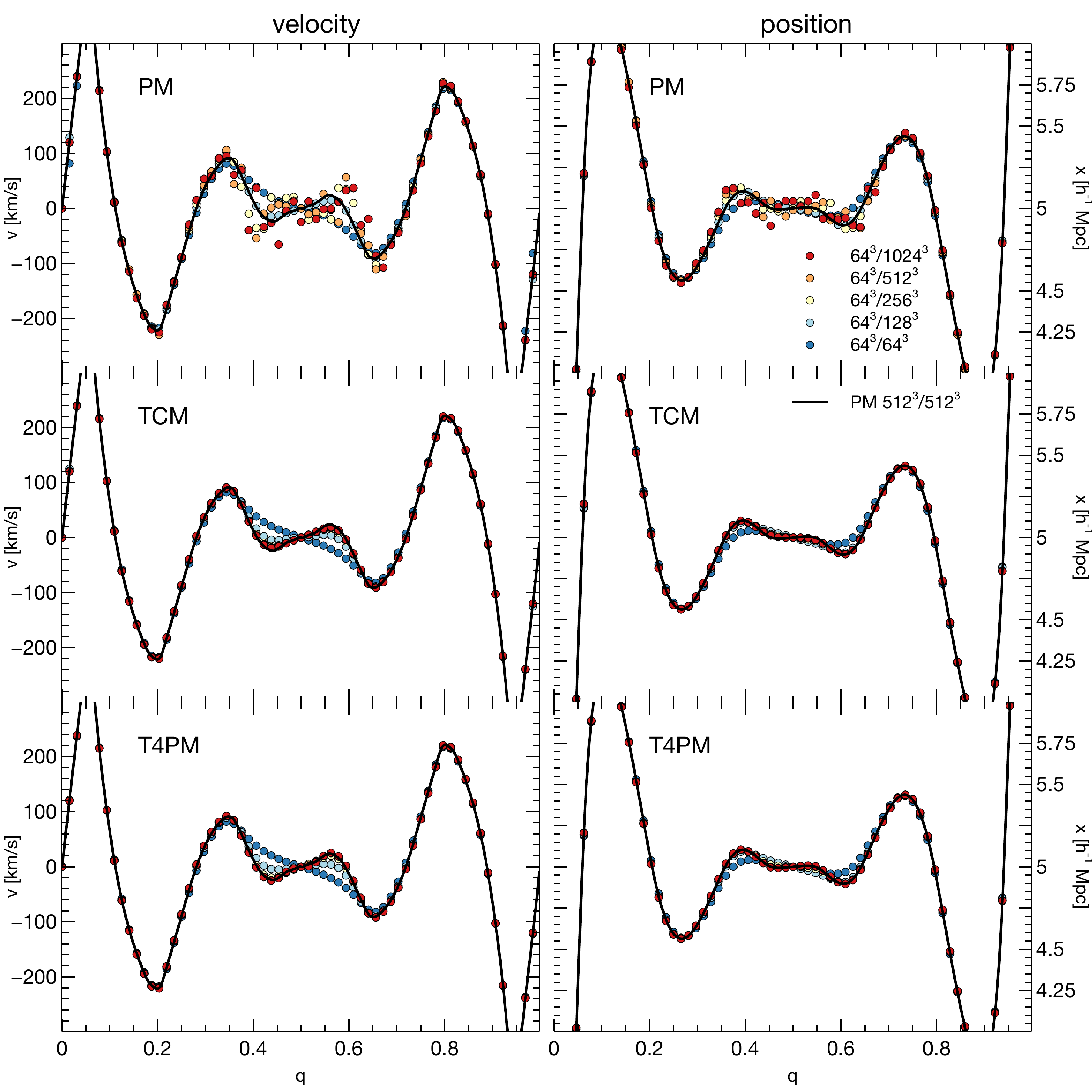}}
\caption{\label{fig:lagrangian_velocity}Velocity (left column) and position (right column)
as a function of the Lagrangian coordinate in the {\em axis parallel} plane wave problem
at the same time shown in Figure \ref{fig:plane_parallel_late}. The top row shows the effect of resolution increase in the
standard PM case, the middle and bottom rows the respective results for the TCM and T4PM case. In all cases the particle number is fixed at $64^3$
and the resolution of the mesh is varied. The black line is the high resolution result for the standard PM method
obtained with $512^3$ particles and a $512^3$ mesh to compute forces.
}
\end{figure}

\begin{figure}
\centerline{\includegraphics[width=0.47\textwidth]{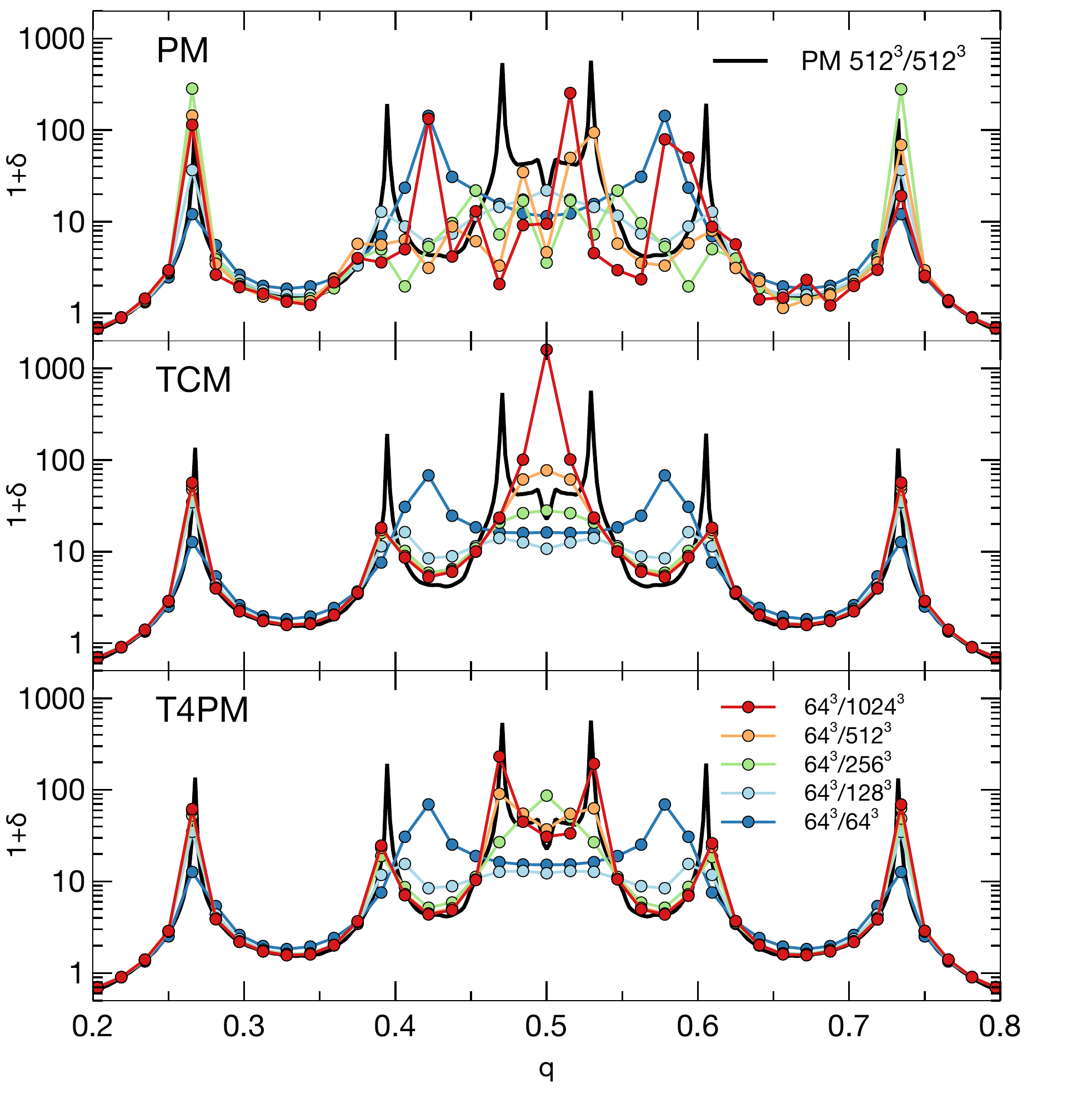}}
\caption{\label{fig:lagrangian_density}Stream density 
as a function of the Lagrangian coordinate in the {\em axis parallel} plane wave problem
at the same time as in Figure \ref{fig:plane_parallel_late}. The top row shows the effect of resolution increase in the
standard PM case, the middle and bottom rows the respective results for the TCM and T4PM case. In all cases the particle number is fixed at $64^3$
and the resolution of the mesh is varied from $64^3$ to $1024^3$. The black line is the high resolution result for the standard PM method
obtained with $512^3$ particles and a $512^3$ mesh to compute forces.
}
\end{figure}

Finally, in Figure~\ref{fig:plane_parallel_late}, we show the phase-space sheet at significantly later times, when $a/a_c=7.7$
and several shell-crossings of the wave have occurred. For a force resolution increasing from $128^3$ to $1024^3$, the 
locations in phase space of the flow tracer particles are shown for TCM and T4PM 
together with the respective locations of the normal $N$-body particles for the standard PM method. They can be compared 
to the solution obtained with standard PM at significantly higher mass
resolution ($512^3$ particles) with a matched force resolution ($512^3$ cells). We clearly see that for standard PM, 
collisionality  destroys the inner spiral structure with increasing force resolution already at a force resolution of $256^3$ cells,
while no such behaviour can be observed for either TCM or T4PM. We also note that for TCM, the innermost part of 
the spiral does not quite approach the correct solution at $512^3$ force resolution, while T4PM perfectly reproduces
the reference solution. 

To illustrate better the errors in particle velocities and positions when the force resolution is varied, we plot
them in Figure~\ref{fig:lagrangian_velocity} as a function of the Lagrangian coordinate $q$ which simply denotes 
the particle position on the dark matter sheet and has been scaled to unit range. We see that collisions in the
case of standard PM affect more strongly the velocities and to a slightly lesser degree the positions. This is
expected as errors in the velocities propagate only averaged over time to the positions. For both TCM and T4PM
we observe convergent behaviour, the solutions approach the reference solution nicely with increasing force 
resolution.

As the most stringent test, in Figure~\ref{fig:lagrangian_density}, we show how the three methods retain the linear approximation to the 
phase space distribution function. To illustrate this, we plot the primordial stream density (see also AHK12),
i.e. the quantity $\delta_i = 2\Delta x / (x_{i+1}-x_{i-1}) - 1$, where $\Delta x$ is the mean particle separation.
This figure impressively highlights how the PM simulation fails to represent the correct evolution of the sheet at high force
resolution. In clear contrast, for TCM we observe 
convergent behaviour: the solution approaches the high-resolution reference solution with increasing force resolution
outside the most central region. Only in the very centre, TCM does not capture the correct density evolution
(consistent with what we see in Figure~\ref{fig:plane_parallel_late}).
For T4PM, we find convergence at all $q$, even the densities in the very centre are followed correctly. We note however
that minor asymmetries occur -- the solution is not perfectly symmetric to $q=0.5$ -- which owes to the asymmetric
decomposition that T4PM employs and originally motivated the use of the symmetric decomposition in TCM.

\subsection{Oblique plane wave collapse}
\label{sec:test_oblique_plane_wave}
We next compare the performance of the new method in the oblique plane wave collapse test problem of \cite{Melott:1997}. This is
again a simple one-dimensional plane wave collapse, but the wave vector is now no longer parallel to any of the cartesian
dimensions. We use the exact 
same set-up as these authors, namely a wave with wave-vector $\mathbf{k}=(2,3,5)k_f$, where $k_f=2\pi/L$ is the fundamental
wave number of the simulation box. First shell crossing occurs at an expansion factor of $1/7.7$ and results are given
at $a=1$ (which is, in fact, also the same set-up as in the axis-parallel case). In contrast to the axis-parallel 
case, now the symmetry of the problem does not match the symmetries of either the initial particle distribution or
the particle mesh.
Using this set-up, \cite{Melott:1997} have observed that any mismatch between force and mass resolution leads to collisionality
that is visible in spurious motion perpendicular to the wave vector and that breaks the symmetry of the solution. Furthermore,
the incongruent symmetries enhance two-body effects, making spurious accelerations much more prominent.

\begin{figure*}
\begin{center}
\includegraphics[width=0.8\textwidth]{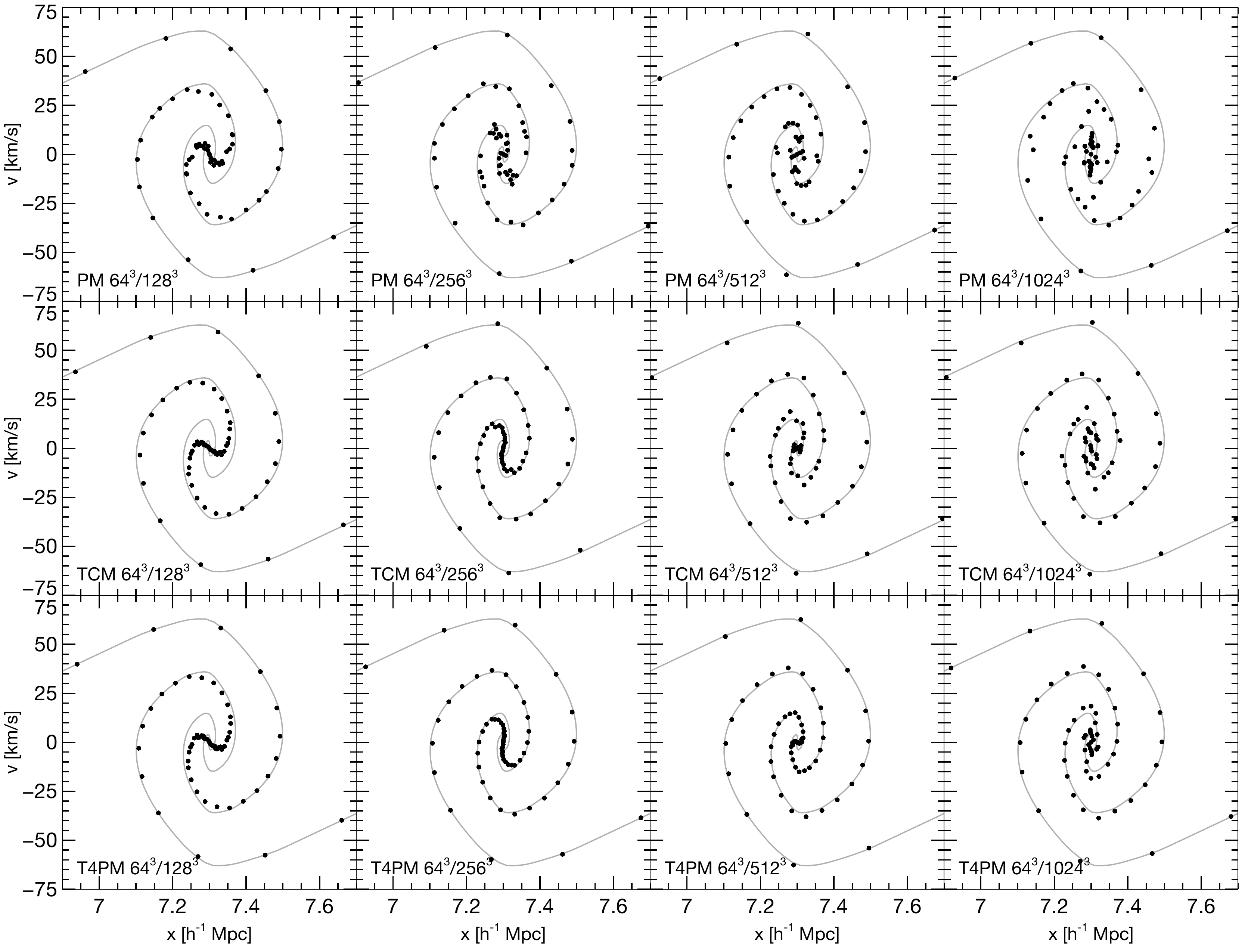}
\end{center}
\caption{\label{fig:oblique_wave}The oblique plane wave collapse problem from Melott et al. (1997) with PM and our new TCM method for $64^3$ particles. The top row correspond
to traditional PM with a mesh resolution of $128^3$, $256^3$, $512^3$ and $1024^3$ cells respectively, while the bottom row correspond to TCM with $128^3$, $256^3$,
$512^3$ and $1024^3$ cells to compute the forces. The panels show the
phase space structure of one sheet at 7.7 expansion factors after the first shell crossing. The grey line is a rescaled version of the high resolution solution obtained for the 
axis parallel plane wave collapse. This is an inherently hard test problem for both codes as its symmetry is neither reflected in the initial particle positions, nor in the
mesh used to compute the forces. }
\end{figure*}

The results of this test are shown in Figure~\ref{fig:oblique_wave}. As discussed by \cite{Melott:1997}, the standard PM solution 
deviates from the correct solution of the spiral (that should match in terms of evolution exactly the one shown in Figure~\ref{fig:plane_parallel_late})
whenever the force resolution exceeds the mass resolution. For the $512^3$ force mesh, the inner part of the spiral in the
standard PM case has disappeared and several  clumpy structures have appeared. In the TCM case, we see no 
strong deviation from the correct solution at $256^3$ force resolution and in the $512^3$ case deviations appear that are however
significantly weaker than those seen for standard PM at $\ge 256^3$. At the highest force resolution we investigated, $1024^3$, we
see a growth of errors that is however restricted to the innermost part of the spiral. As expected, T4PM clearly outperforms both 
standard PM and TCM. The solution appears smooth with no scattered particles up to $512^3$ force resolution, and shows only
small errors in the very innermost region of the spiral at $1024^3$. For both TCM and T4PM, no significant errors
occur in the outer regions of the spiral at all force resolutions, while for standard PM errors are less well confined to the innermost
region.

We note however that this is an inherently hard problem for all methods as the symmetry of the problem is neither reflected in the 
initial conditions, nor in the methods, so that deviations from that symmetry are expected to arise easily and will affect
the solution quickly.  

\subsection{Antisymmetrically perturbed plane wave collapse}
\label{sec:test_ripple}

\begin{figure*}
\begin{center}
\includegraphics[height=0.95\textwidth,angle=270]{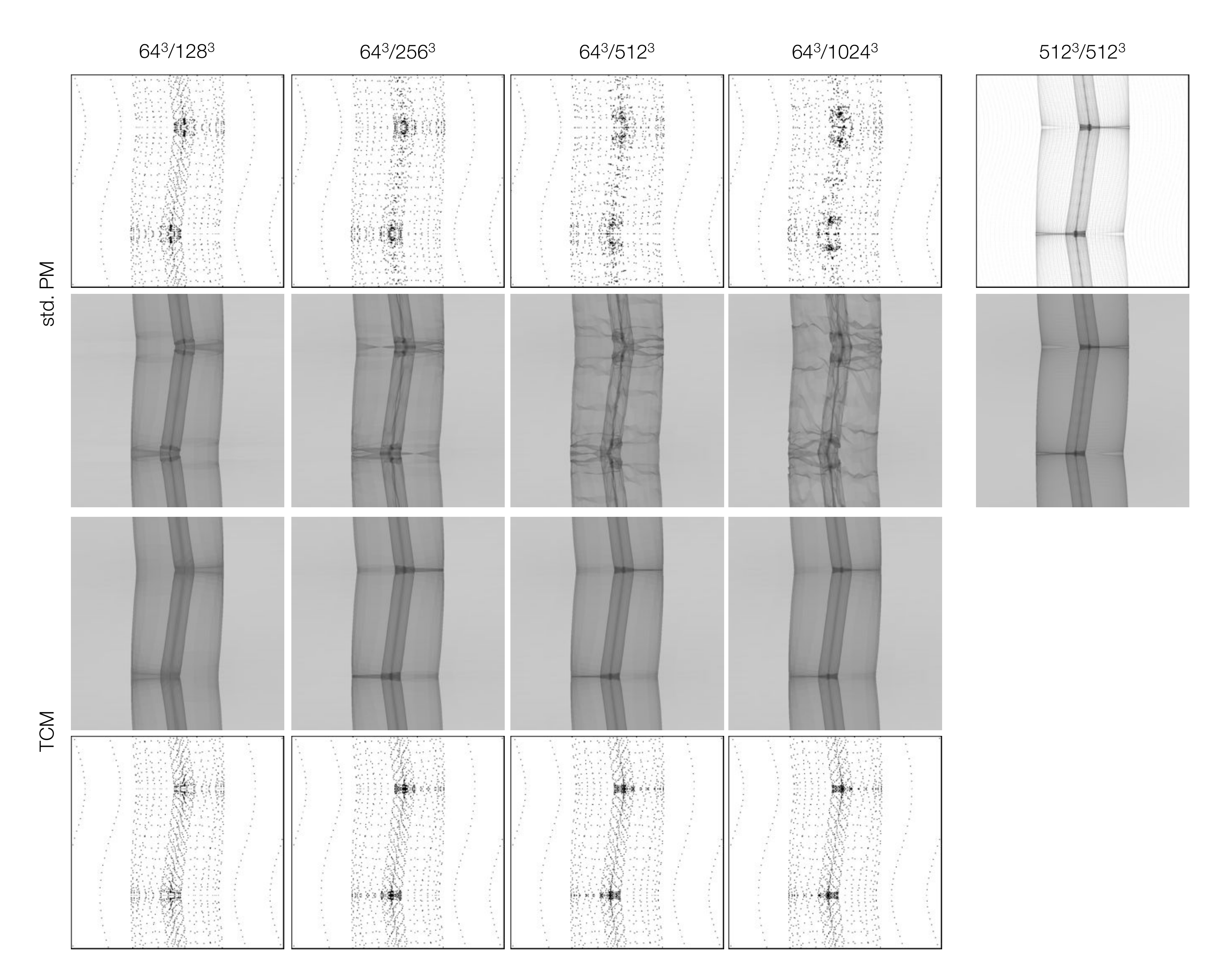}
\end{center}
\caption{\label{fig:ripple_wave_008}Collapse of an antisymmetrically perturbed plane wave. The panels give a zoom-in, shortly after
shell-crossing occurred also in the y-direction. Shown are the particle positions at $a/a_c=4.6$. The two left columns show results obtained with
the new TCM method, the right columns the results obtained with the standard PM method. In all cases, $64^3$ particles were used and the
resolution of the force mesh was varied from $128^3$ to $256^3$ to $512^3$ to $1024^3$ (top to bottom respectively).  The inner columns show the 
projected density computed using the dark matter sheet method, the outer columns the tracer particle positions. The row at the very bottom shows the reference
result obtained with the standard PM method with $512^3$ particles and a matched force resolution of $512^3$.
Two-body effects destroy most of the caustic structure when standard PM is used with high force and low mass resolution. 
}
\end{figure*}

\begin{figure*}
\begin{center}
\includegraphics[height=0.95\textwidth,angle=270]{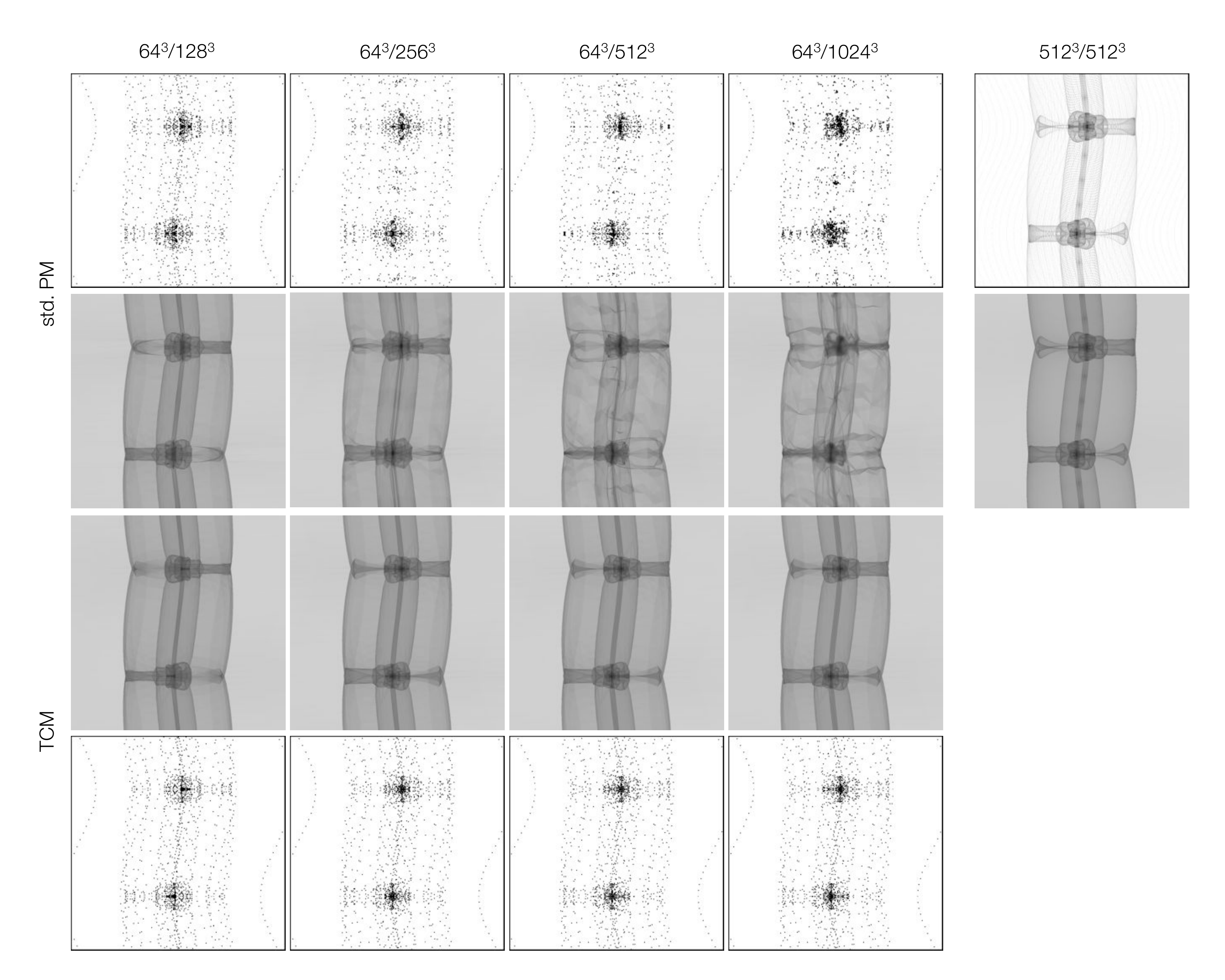}
\end{center}
\caption{\label{fig:ripple_wave_013}Same as Figure \ref{fig:ripple_wave_008} but at a later time $a/a_c=7.7$, i.e. $a=1$.}
\end{figure*}

In addition to the one-dimensional problems considered above, we now turn to a two-dimensional test problem.
For this, we consider the antisymmetrically perturbed wave described by \cite{Valinia:1997}. This is
a two dimensional problem in which the initial gravitational potential is given by that of a plane wave in x-dimension 
with a sinusoidal phase perturbation in the y-dimension
\begin{equation}
\phi(\mathbf{x}) = \bar{\phi} \cos\left(k_p \left[ x + \epsilon_a \frac{k_p}{k_a^2} \cos k_a y \right]\right).
\end{equation}
The initial particle positions and velocities are again obtained using the Zel'dovich approximation. We adopt 
$k_p = 2\pi / L$, $k_a = 4\pi/L$ and $\epsilon_a=0.2$, where $L=10\,h^{-1}{\rm Mpc}$ is the size of the simulation box.
Also, $\bar{\phi}$ is again chosen so that first shell crossing occurs as in Section \ref{sec:test_oblique_plane_wave} 
at an expansion factor of $a_c=1/7.7\simeq0.13$. The simulation
is evolved to $a=1$. We show the results for $64^3$ particles using the TCM and standard PM methods for various force
resolutions at $a=0.6$ in Figure \ref{fig:ripple_wave_008} and at $a=1$ in Figure \ref{fig:ripple_wave_013}\footnote{These
Figures correspond to the same region shown by \cite{Valinia:1997} in their Figure 6. Figure
\ref{fig:ripple_wave_008} corresponds to a time shortly after their panel 6d, while Figure \ref{fig:ripple_wave_013}
corresponds to a time shortly after their panel 6e. These authors used a standard PM method in two dimensions.}. 

In Figure   \ref{fig:ripple_wave_008}, we show the particle positions and projections of the full tetrahedra 
(using the method described in AHK12), for both standard PM and TCM at a time shortly after shell-crossing happened also along
the y-direction at $a/a_c=4.6$. For comparison, we also show results obtained using standard PM with significantly higher mass resolution
$512^3$ and a matched force resolution of $512^3$ cells. We do not show the results for T4PM in that figure as they are
qualitatively identical to those for TCM.
For standard PM, increasing the force resolution leads to two-body effects that destroy most of the structure
of the caustics. The inner vertical caustic has virtually disappeared, instead clumps of particles emerge. In the tetrahedron
projection images, these effects appear as increasing levels of small scale noise. The two locations
in the images where shell-crossing has occurred along two dimensions show no sign of convergence with increasing force
resolution. Rather than becoming narrower features with increasing force resolution, in std. PM, they instead appear more 
and more extended and dissociate into two separate clumps at the highest force resolutions considered . In contrast to this,
for TCM we observe convergence, all caustic features are visible at all force resolutions and only become narrower when
the force resolution is increased. The agreement between TCM and the high mass-resolution PM result is
remarkable, as is the lack of convergence of the low-mass resolution PM result.

In Figure~\ref{fig:ripple_wave_013}, we show results at much later times $a=1$. 
Now, exactly two large clumps should have formed which are correctly produced
by both methods at all force resolutions. Again, we observe converged caustic structures for TCM, while they 
undergo severe distortion with increasing force resolution for standard PM. 
For comparison, we again present also the high mass resolution result for standard PM. 
Interestingly, at the highest force resolution in the PM result, a small dense clump of particles appears between
the two genuine clumps: clearly an artificial fragment. When rendering the tetrahedra, the
clump almost disappears indicating that it cannot correspond to a region of convergent flow.
In general, at the highest force resolution we consider, the PM result bears little resemblance
with the reference solution. It is clear that this would only worsen if the force resolution were
increased even further. Most remarkably, in contradiction to the reference case, the large clump 
forms by a merger of the smaller clumps that are visible in Figure~\ref{fig:ripple_wave_008}.
In TCM we observe no such fragmentation and subsequent merging of the fragments even
at the highest force resolution considered. Again, at the scale of the images shown in the figure,
the results for T4PM are qualitatively identical to those for TCM, only the detailed caustic structure
in the large clumps shows some differences, so that we investigate those in more detail now.

\begin{figure}
\begin{center}
\includegraphics[width=0.3\textwidth]{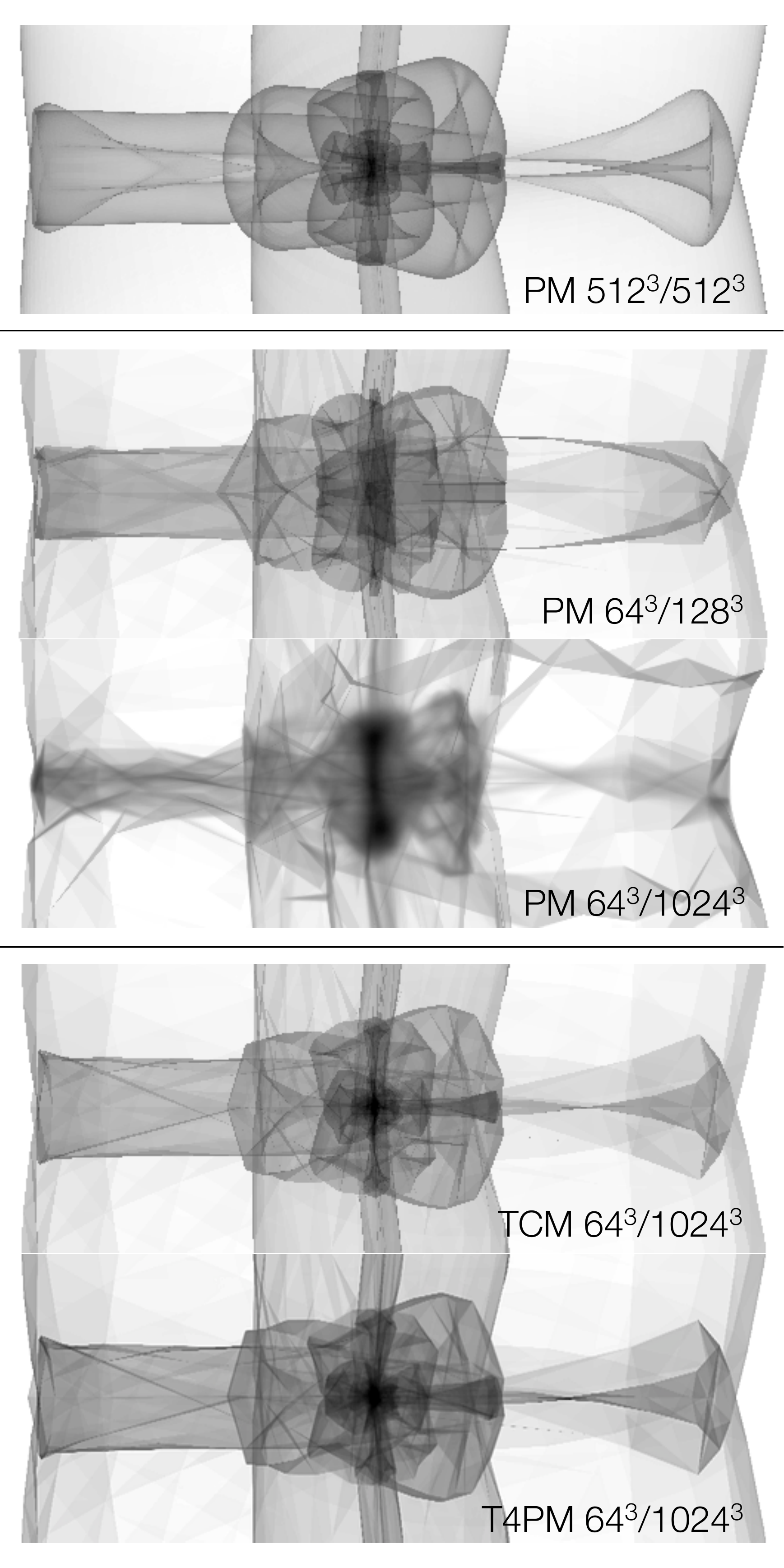}
\end{center}
\caption{\label{fig:ripple_detail}Detail of the caustic structure from Figure~\ref{fig:ripple_wave_013} for a selection of
resolutions. The top panel shows the reference
solution obtained with the standard PM method with $512^3$ particles and $512^3$ force resolution.
The second and third panels from the top show results for the standard PM method with $64^{3}$ particles and $128^{3}$ 
and $1024^3$ force mesh resolution, respectively. High force resolution further combined with low
mass resolution leads to very strong particle-particle scattering in std. PM (cf. Figure~\ref{fig:ripple_wave_013}).
The fourth and fifth
panel show results obtained with the new TCM and T4PM methods, respectively,  using $64^{3}$ flow tracers and 
$1024^3$ force resolution.}
\end{figure}

In Figure~\ref{fig:ripple_detail}, we show the detailed structure of the clumps from Figure~\ref{fig:ripple_wave_013}
for a few select cases of force resolution. The top panel shows the reference PM solution to which we compare
the standard PM solutions at a force resolution of only $128^3$ (middle left) and $256^3$ cells (bottom left), as 
well as the TCM (middle right) and the T4PM (bottom right) result at $1024^3$ force resolution.
We note that at relatively low force resolution of $128^3$ cells, the standard PM method reproduces larger scale caustics
correctly, but, as expected, the central density is low at low force resolution, and the innermost caustics are not
coincident with those in the reference case. Increasing the force resolution does not approach the correct solution
however. With the $1024^3$ force mesh, we observe a numerical solution that bears almost no resemblance 
to the reference solution.
We note that both TCM and T4PM are able to reproduce both the small-scale and large-scale caustic structure
in very good agreement with the reference case, using only $64^3$ instead of $512^3$ flow tracers. 
At the same time, a high central density is reached. In fact, this density is somewhat too high. 
This is a sign of a breaking down of the linear approximation as discussed
in Section~\ref{sec:linearerrors}. The phase-space sheet has undergone super-Lagrangian growth in this 
region and refinement of our decomposition would be needed to accurately follow its evolution. Noting
this limitation, we want to emphasise however how well the linear approximation represents the
caustic structures and even small details of the density field outside of these heavily mixed regions.
The problem is also present but slightly less severe for T4PM. We also note that the position of the outermost 
caustic (measured along the vertical direction) is closer to the centre for TCM
than for T4PM and the reference case. This is also related to the loss of resolution with the consequential errors
in mass deposition.


\section{Three-dimensional hot dark matter simulations}
To complete our first investigation of the virtues and limitations of the  TCM and T4PM
methods, we next consider the case of three-dimensional cosmological structure formation from random
initial conditions with an initial power spectrum with a well resolved 
cut-off scale. Such simulations are well known to exhibit artificial fragmentation
in standard $N$-body methods.

\label{sec:HDM-simulations}
\subsection{Specifics of the Simulations}

We have carried out cosmological $N$-body simulations of a volume of
$40\,h^{-1}{\rm Mpc}$ length.  The initial conditions for these
single-mass-resolution simulations were generated with the {\sc Music}
code \citep{Hahn:2011} keeping large-scale phases identical with
changing mass and spatial resolution. We assume a toy
$\Lambda$HDM cosmological model with density parameters $\Omega_{\rm
  m}=0.276$, $\Omega_{\Lambda}=0.724$, power spectrum normalization
$\sigma_8=0.811$, Hubble constant $H_0=100\,h\,{\rm km}{\rm
  s}^{-1}{\rm Mpc}^{-1}$ with $h=0.703$ and a spectral index
$n_s=0.961$. 
The transfer function translating the primordial spectrum of density
fluctuations into the post-recombination era assumes a  300$\,{\rm eV}$ 
dark matter particle,  appropriate for a hot dark matter fluid with a 
free streaming scale of
$\sim7.7\,h^{-1}\,$Mpc, i.e. approximately one fifth of the box size we
simulate. While being an utterly unrealistic model for our own Universe,
this ensures plenty of resolution in the sheets and
filaments expected to form before any haloes originate on that scale
and will give us an opportunity to detail both the shortcomings of the
old approaches and the benefits of the new method. All the simulations
are started at redshift one hundred, well in the linear regime, and evolved
down to $z=0$. 

\subsection{A remark on fluids with a finite temperature}
We note here that our novel method in its current state only applies
to perfectly cold collisionless fluids where the dark matter sheet is infinitely
thin. One might thus wonder if it can accurately 
model fluids which have a non-zero temperature such as WDM or HDM fluids. 
For the $300\,{\rm eV}$ toy problem that we study here, the microscopic velocity
dispersion at $z=0$ is in fact $\sigma_v\simeq0.22\,{\rm km/s}$ and scales 
with the expansion factor $a^{-1}$ to earlier times. Its impact on the
initial conditions is adequately captured in the transfer function by erasing all
perturbations below the maximum free-streaming scale. We note
that this velocity dispersion describes the {\em microscopic} random velocities.
Any attempt to sample the microscopic dispersion with macroscopic
resolution elements will yield inconsistently large bulk motions, the finite thickness
of the sheet would thus have to be modelled directly by incorporating it in the
discretisation of the distribution function. 
Modelling a genuinely six-dimensional initial phase-space distribution would require
a six-dimensional extension of our method (which is beyond the scope
of this first paper). At the same time, the mean evolution is expected to
be still well described by the perfectly cold limit. Furthermore, even for such a 
light DM particle, the velocity dispersion is at late times significantly 
lower than typical velocities that arise during gravitational collapse, such that
differences away from the centers of haloes are expected to be small.

We wish to remark here also
that it is precisely in the cold limit in which standard $N$-body methods have shown
their shortcomings: a cold fluid with a finite perturbation spectrum (i.e. with no
perturbations below some scale) is in principle unstable to all perturbations
since the Jeans length is effectively zero everywhere. If any perturbations are
introduced due to the discretization scheme, they can and will grow. We will
show below that such behaviour is much reduced in our new method.

\subsection{Analysis I : force and density fields}

\begin{figure*}
\begin{center}
\includegraphics[width=0.75\textwidth]{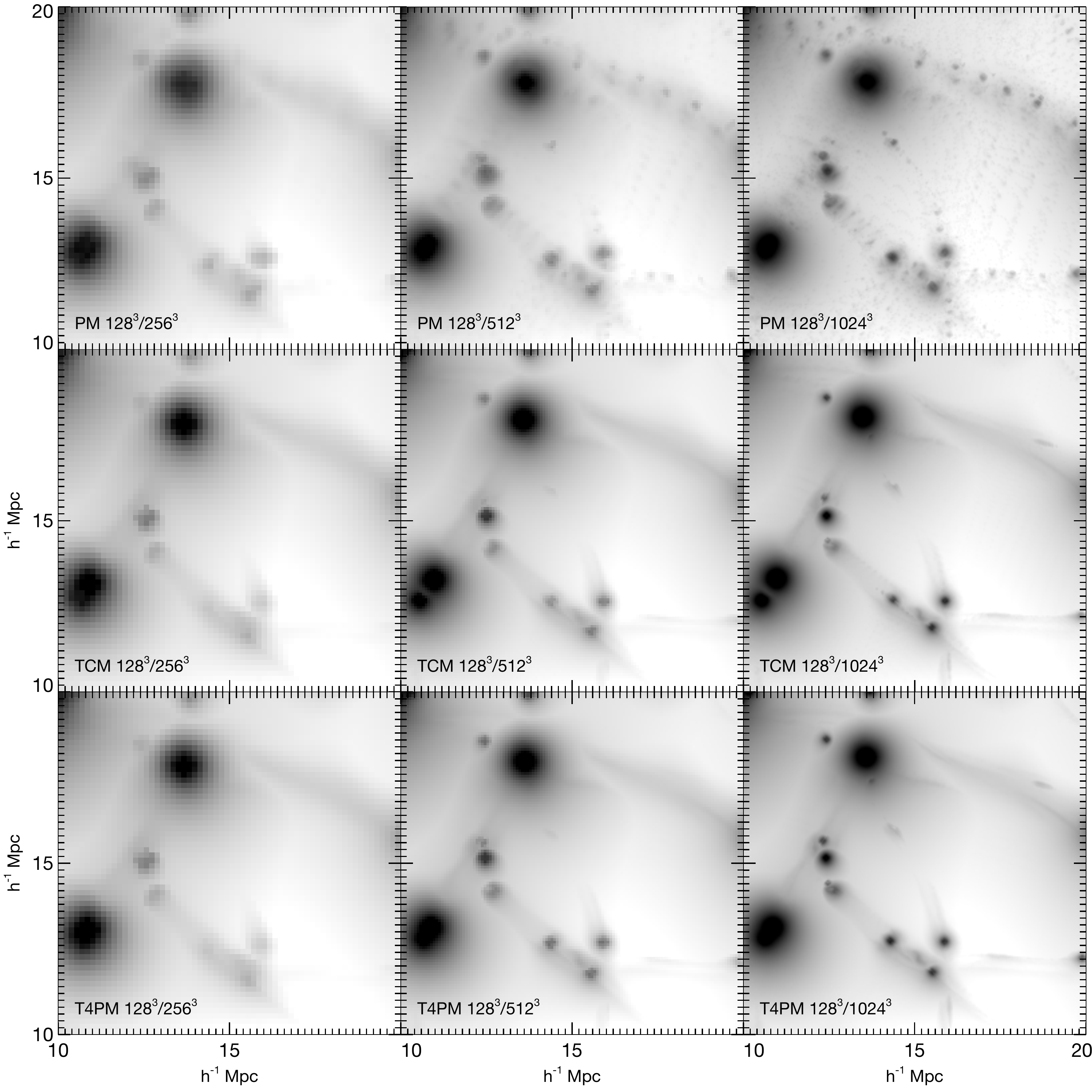}
\end{center}
\caption{Maximum intensity projections of the magnitude of the gravitational force, $\log_{10}\| \boldsymbol{\nabla} \phi\| $, in a region
$10\times10\times40\,h^{-3}{\rm Mpc}^3$ for simulations with fixed mass and varying force resolution. The image resolution directly
corresponds to the resolution of the mesh used to compute the gravitational force in the simulations. The force resolution increases from
$256^3$ cells (left column) to $512^3$ cells (center) to $1024^3$ (right column). (Top row:) results for the standard PM method with $128^3$ particles. 
(Middle row:) results for the TCM method, using $128^3$ tracer particles. (Bottom row:) results for the T4PM method with $128^3$ tracers. 
The artificial fragmentation of the filaments is clearly visible for the standard PM method at high force resolution. 
No such fragmentation is visible for neither TCM nor T4PM.}\label{fig:forcemaps}
\end{figure*}

\begin{figure*}
\begin{center}
\includegraphics[width=0.75\textwidth]{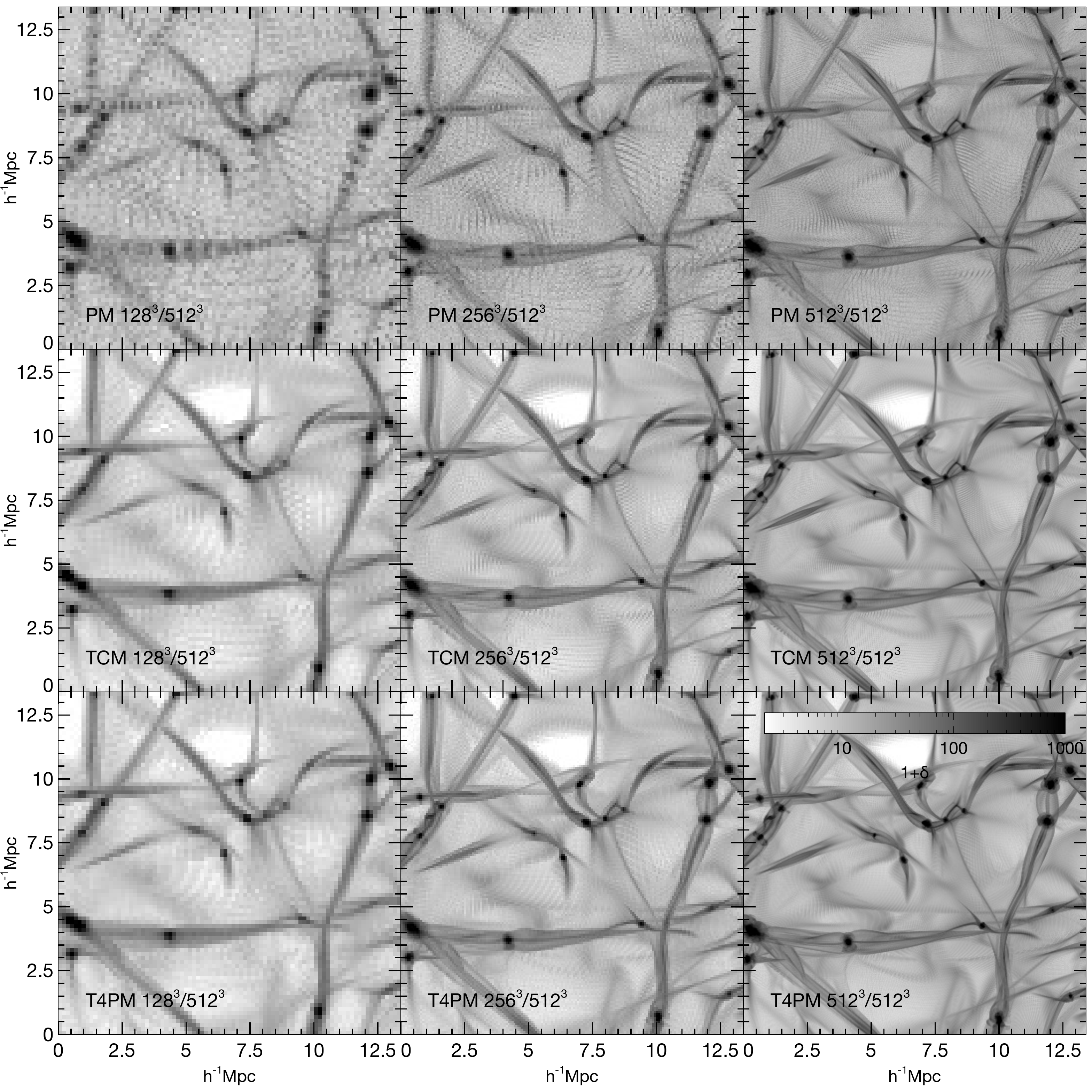}
\end{center}
\caption{Density fields at fixed force resolution with increasing mass resolution for the two methods.
Shown is the maximum density projection of a $13\times13\times 40
  h^{-3}{\rm Mpc}^3$ region in the $128^3$, $256^3$ and $512^3$ particle
  simulations for standard PM (top row), TCM (middle row) and T4PM (bottom row).
  In all cases a $512^3$ mesh was used to compute the forces.
  CIC density estimates were obtained on a mesh with  two times the number 
  of cells compared to the number of particles per linear dimension, i.e.
  $256^3$ for the $128^3$ particle runs for example.  Compared to
  the PM simulations, all the beads-on-a-string artificial fragments are
  gone in TCM and T4PM. }\label{fig:densityprojection}
\end{figure*}

In Figure~\ref{fig:forcemaps}, we show maximum intensity projections
of the magnitude of the gravitational force at $z=0$ for the 
standard PM method (top row), TCM (middle row) and T4PM (bottom row). We keep the mass resolution constant
at $128^3$ particles and flow tracers, respectively, and vary the force resolution
between $256^3$ (left column), $512^3$ (middle column) and $1024^3$ cells (right column).
The images directly show the magnitude of the force $\|\boldsymbol{\nabla}\phi\|$
on the actual mesh that is used to compute the forces in the simulation,
i.e. each pixel matches a force resolution element.
We clearly observe fragmentation of the filamentary regions for 
the standard PM method when the force resolution is $512^3$ or
$1024^3$. Note that even $1024^3$ corresponds only to a factor
of 8 with respect to the mass resolution, still a factor of 4-8 short
of the resolution typically employed for cosmological simulations.
For TCM and T4PM, we see no sign of fragmentation of the filaments,
instead they appear perfectly smooth, and even the associated
caustics become visible in the force.
At the same time, the limitations of our new methods become clearly visible.
The force in the haloes is significantly larger,
again reflecting the fact that our linear approximation to the distribution
function breaks down there. This is further complicated by the zeroth
order mass assignment scheme that we adopted for TCM, where mass is
deposited in the centroid position only. If the tetrahedra are well
mixed inside the haloes, the centroid is more likely to lie at the center position of
the halo, leading to a bias in the gravitational force. This error can propagate
to rather large scales, as can be seen from the large halo in the lower left
side of the images. Only one halo is visible for standard PM while TCM predicts
two distinct structures at higher force resolutions. 

The additional bias in TCM due to all mass being deposited close to the
center of a halo is of course remedied in T4PM which shows no such large-scale
discrepancies to standard PM in the positions of haloes. The remaining error
must thus be dominated by errors due to the piecewise linear approximation
to the distribution function in strongly mixed regions.This can only be circumvented
by refinement techniques (see also our discussion in Section~\ref{sec:linearerrors}).

While our method, without refinement, does not capture the correct
dynamics in the high density regions of haloes, it clearly performs
significantly better in regions of moderate overdensity compared
to the standard PM method.

We next consider the case of fixed force and varying mass resolution.
In Figure~\ref{fig:densityprojection}, we show maximum intensity 
projections of the CIC density field for standard PM (top row), TCM (middle row)
and T4PM (bottom row). In all cases, we used a $512^3$ mesh to
compute the forces. We use $128^3$ (left), $256^3$ (middle)
and $512^3$ (right) particles and flow tracers, respectively.
Our findings are much in line with the results above. At a mass
resolution $1/4$ the force resolution, the PM method shows
strong fragmentation of the filaments that becomes weaker when
the mass resolution is increased and disappears entirely
when force and mass resolution are matched. At the same time,
with TCM and T4PM, we see no sign of fragmentation.
As before, we observe increased densities inside haloes in TCM,
as well as errors in the positions of the most massive
haloes for TCM, both are much smaller with T4PM. At the same time, we obtain a perfectly
converged density field in low and intermediate density regions.
In contrast to the PM results, all features of the density field present 
at low mass resolution are present also at higher resolutions. 

The lack of artificial fragmentation of the new methods allows to measure the abundance
of collapsed structures reliably. We show in \cite{Angulo:2013} that T4PM yields
a reliable mass function of haloes in WDM cosmologies with virtually no artificial
haloes while Tree-PM produces the expected large fraction of small scale clumps. 

\subsection{Analysis II : density probability distribution functions}

\begin{figure}
\begin{center}
\includegraphics[width=0.4\textwidth]{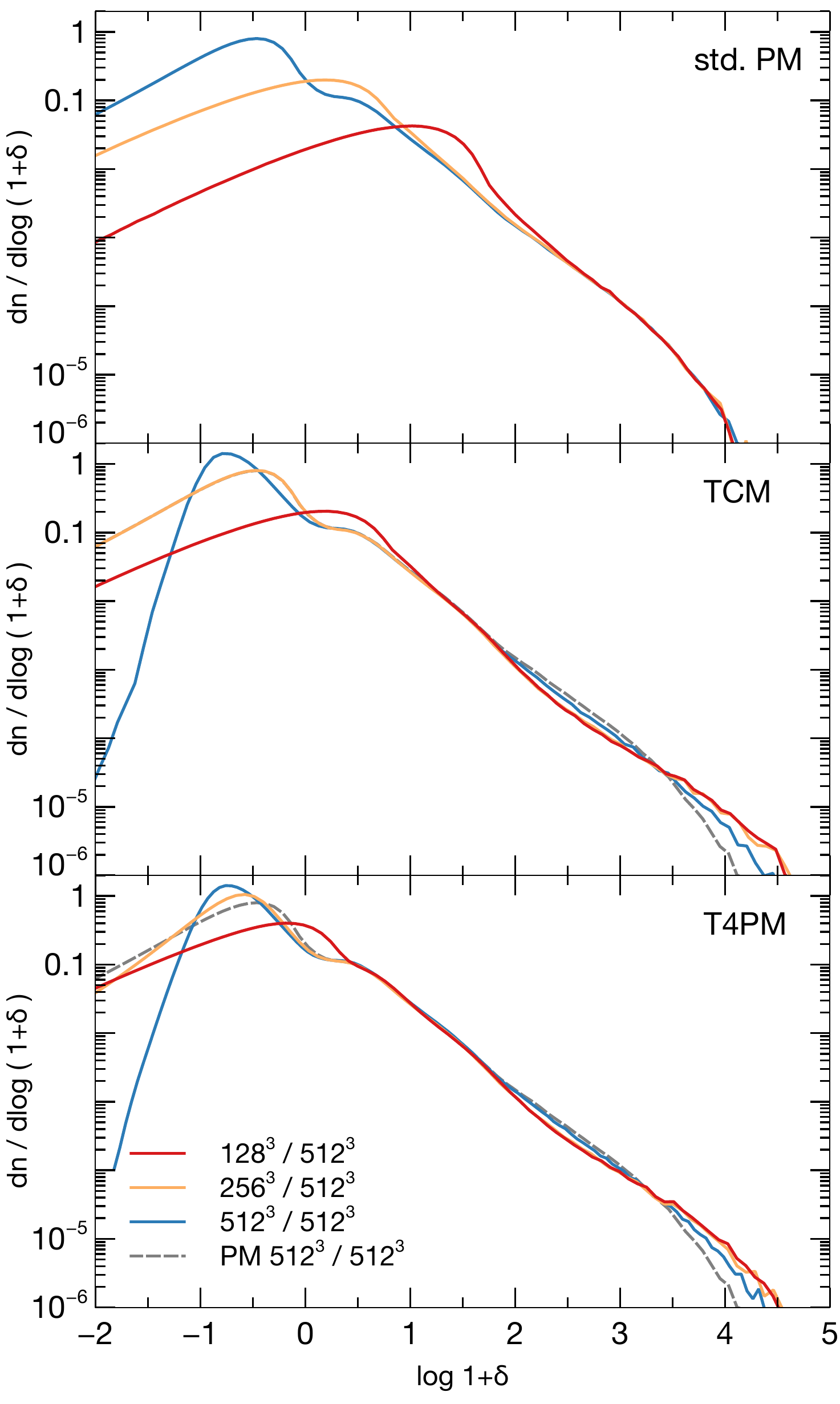}
\end{center}
\caption{\label{fig:density_pdfs} Probability distribution functions of the CIC density field for the HDM simulations at fixed
force and varying mass resolution. All PDFs were obtained from a $512^3$ mesh, i.e. the density fields 
are the exact density fields that source the gravitational forces in the gravity solver in each case.
}
\end{figure}

\begin{figure}
\begin{center}
\includegraphics[width=0.4\textwidth]{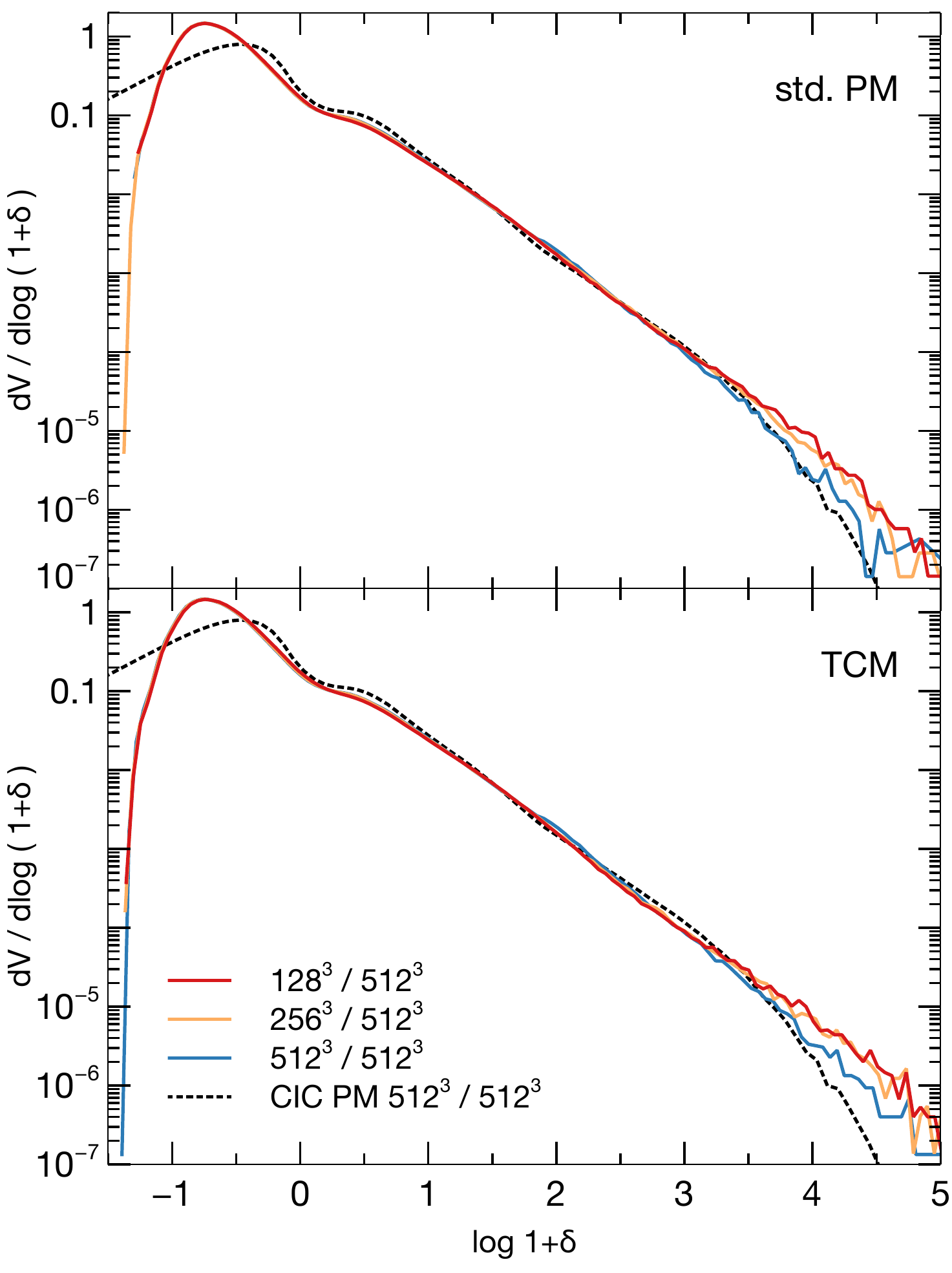}
\end{center}
\caption{\label{fig:density_pdf_sheet}Probability distribution functions of the density computed using the dark matter sheet tesselation for
the HDM simulations at fixed force and varying mass resolution. Shown are the density estimates for the standard PM method (top)
as well as for TCM (bottom) using $128^3$ (red), $256^3$ (yellow) and $512^3$ (blue) particles. For comparison, the density PDF
obtained using CIC for the standard PM method from Figure~\ref{fig:density_pdfs} is also shown (black dotted).
}
\end{figure}

We now complement our visual analysis of the $\Lambda$HDM simulations with
density probability distribution functions (PDFs). First, in Figure~\ref{fig:density_pdfs}, we
show the density PDFs obtained using the CIC mass deposition and thus from the
density fields that are used as the source terms in the force calculation of the simulations.
In all cases, for standard PM as well as for TCM and T4PM, we observe a slow convergence
of the intermediate to low density part of the PDF with increasing particle number. This
reflects exactly what we stressed already in the Introduction, namely that many particles 
are needed to obtain a reasonable density estimate in any point. At the lowest resolution
we considered, i.e. $128^3$ particles, the density PDFs for standard PM is only converged at overdensities
$\delta\gtrsim 100$. The new methods TCM and T4PM  benefit from the increased number
of mass tracing particles so that the density PDF is converged at $\delta\gtrsim10$ in the
case of TCM and $\delta\gtrsim2$ in the case of T4PM when using $128^3$ flow tracing particles. 
With the same number of flow tracers as in the standard PM case, we thus achieve a much
improved density estimate at intermediate densities. This clearly indicates that at these densities,
the number of flow tracers is not the limiting factor of a good density estimate, but the CIC
mass assignment.
While TCM and T4PM thus perform significantly better at low and intermediate densities, 
we clearly see the limitation of these methods that we discussed before. Both, TCM and
T4PM do not converge at the highest densities. We see here clear evidence that
at densities $\delta\gtrsim 100$, densities are consistently shifted upwards compared
to the standard PM method with a pronounced excess at the highest densities $\delta\gtrsim 2000$.
As discussed before, we expect that the methods break down in regions of heavy mixing
which should only occur inside of haloes, consistent with the deviations that we see only
at densities $\delta\gtrsim 100$.

Finally,  we also investigate the density PDFs obtained from density estimates that 
directly use the tesselation of the dark matter sheet density (as in AHK11). 
The results for standard PM
and TCM are shown in Figure~\ref{fig:density_pdf_sheet}.  We first observe that 
we obtain density PDFs that are completely independent of resolution at
densities $\delta\lesssim 1000$ demonstrating the much improved quality of density estimates
based on the dark matter sheet when compared to the CIC estimates. At the same time,
we now see that all estimates based on either standard PM or TCM do not converge
at the highest densities for the resolutions we considered. Since the CIC estimate for
standard PM was perfectly converged at those densities, we can conclude that
it is in fact the piecewise linear approximation that breaks down in the highest density
regions. Since the density estimate is perfect in all other regions even with the
lowest number of flow tracers considered, it is clear that the resolution needs to be
improved only in regions of heavy mixing and not globally.

\subsection{Analysis III : density power spectra}

We finally quantify the differences in the density fields sourcing the gravitational
forces using density power spectra. We calculate these
power--spectra from the actual density field that is used in the gravitational force
determination. We consider here the case of a force resolution fixed to $512^3$
cells and varying mass resolution from $128^3$ to $256^3$ to $512^3$ particles
and flow tracers, respectively, i.e. for the same simulations as in Figure~\ref{fig:densityprojection}.
The bin positions and sizes are identical for all spectra we show. 

\begin{figure*}
\begin{center}
\includegraphics[width=0.9\textwidth]{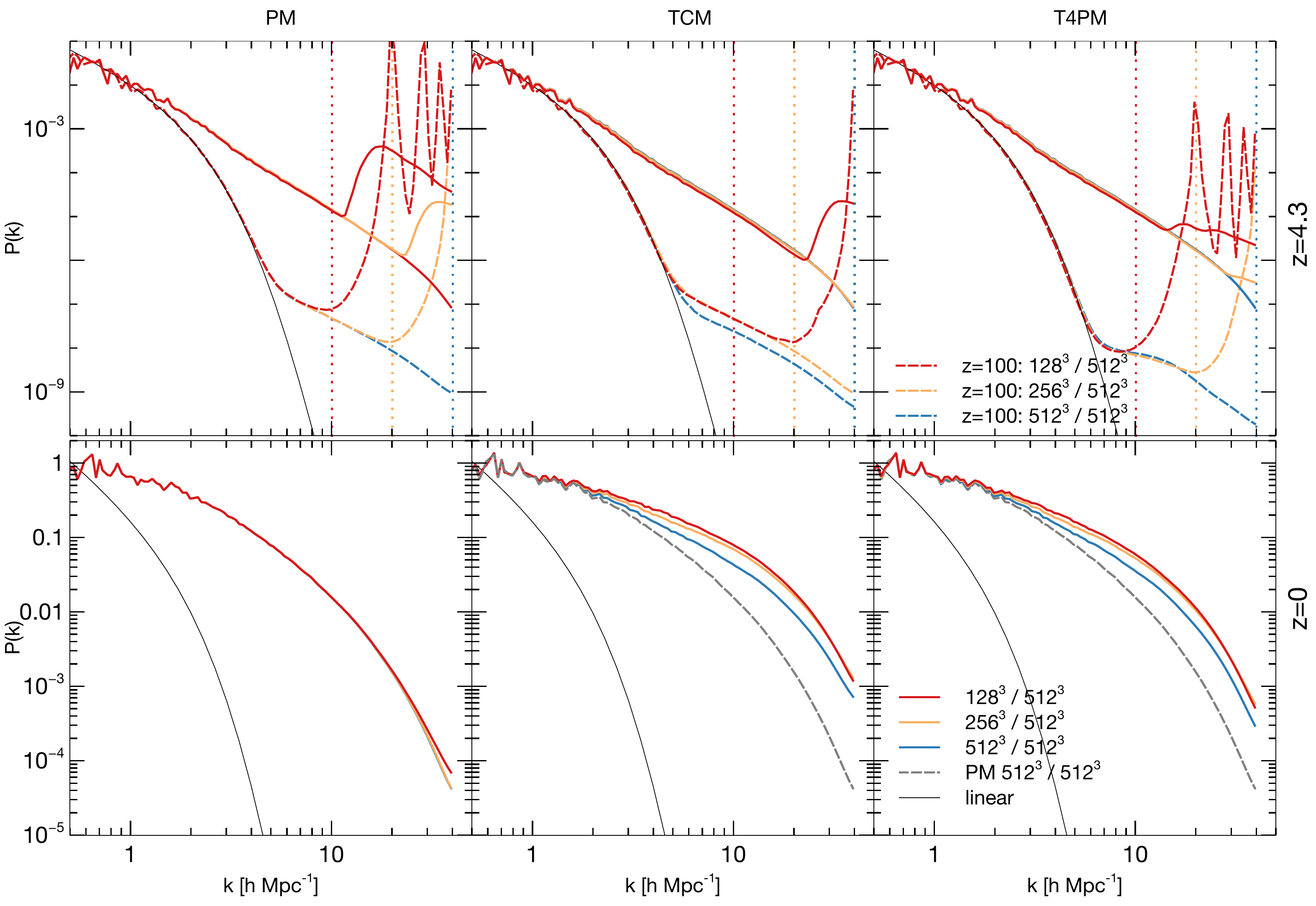}
\end{center}
\caption{\label{fig:power_evol}The density power--spectra at three redshifts, $z=100$ and $z=4.3$ (top row panels),
as well as $z=0$ (bottom row panels) for our resolution study where the particle number was changed from 
$128^3$ (red) to $256^3$ (yellow) to $512^3$ (blue)
and the force resolution was kept constant at $512^3$ cells. The power--spectra shown are computed using the same mass assignment
and mesh resolution as in the actual simulation (left panels: standard PM, centre panels: TCM, right panels: T4PM). For comparison,
we show the reference result obtained with standard PM using $512^3$ particles and a force resolution of $512^3$ as a grey dashed
line, as well as the linearly scaled initial power spectrum as a thin black line. In the top row panels, the initial power spectra at $z=100$
are also shown by dashed lines. The coloured vertical dashed lines in the top panels indicate the Nyquist wavenumber of the 
$128^3$, $256^3$ and $512^3$ $N$-body and tracer particles respectively. While the new methods
perform well in this test at high redshift, the strong mixing in the interior of haloes at late times leads to a runaway overestimation of the
density inside haloes that is stronger in TCM than in T4PM. At the same time, it is remarkable how none of the problems of standard
PM (cf. Figures \ref{fig:forcemaps} and \ref{fig:densityprojection}) stands out in this commonly used diagnostic of cosmological simulations.
}
\end{figure*}

\begin{figure}
\begin{center}
\includegraphics[width=0.4\textwidth]{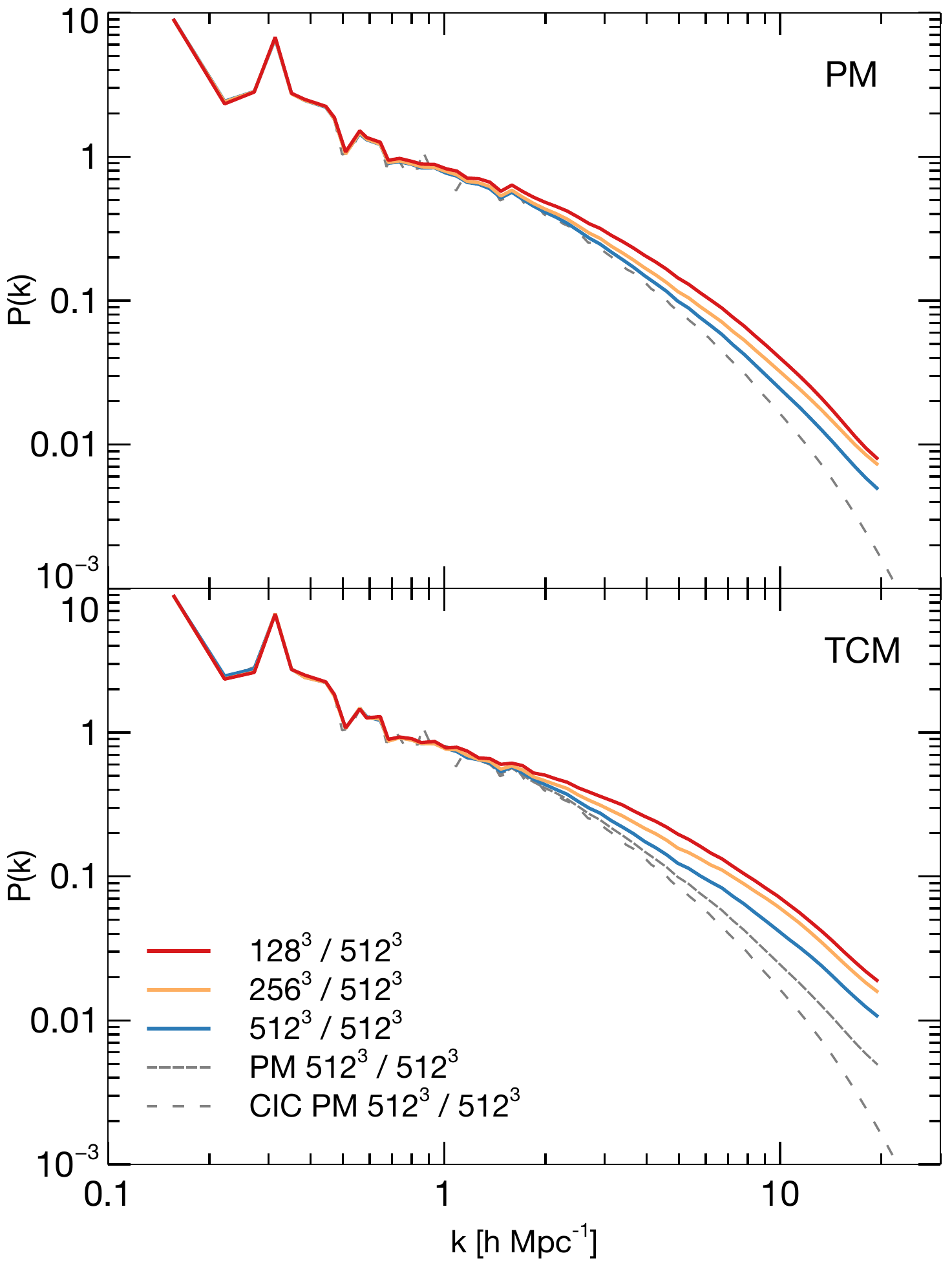}
\end{center}
\caption{\label{fig:power_sheet}The density power--spectra at $z=0$ computed from the dark matter sheet tesselation for
the HDM simulations at fixed force and varying mass resolution. The tesselation has been projected onto a $256^3$ mesh
from which the power spectrum is computed via FFT. The top panel shows the results for standard PM, the bottom panel
the results for TCM using $128^3$ (red), $256^3$ (yellow) and $512^3$ particles (blue). In the bottom panel, for comparison,
we also show the results for the standard PM method with $512^3$ particles using both the tesselation and simple CIC
deposit into a $512^3$ mesh (same curve as in Figure~\ref{fig:power_evol}, bottom left panel).
}
\end{figure}

Figure~\ref{fig:power_evol} shows the density power spectra at two times,
$z=4.3$ (top row) and $z=0$ (bottom row). Already in the high redshift
spectrum, significant non-linear growth has occurred (the linearly scaled
initial power spectrum is shown in the figure as a thin black line).
At these early times, we see
that all methods converge to the same result, albeit with important differences
in their convergence behaviour:
\begin{itemize}
\item Standard PM: the power spectra at wave numbers smaller than the Nyquist wave number of the
initial particle lattice agree perfectly, while a pronounced peak is visible at the Nyquist
wave number that simply shifts to larger $k$ with increasing mass resolution.
\item TCM: As in standard PM, a prominent peak is visible that is however shifted to larger $k$,
as expected because the mass tracers have a two times larger wave number in the initial
particle lattice than the flow tracers. Also, we observe that solution at low mass resolution
very slightly undershoots the solution at higher mass resolution at intermediate wave numbers.
\item T4PM: Quite in contrast to both standard PM and TCM, no pronounced peak at large
wave numbers is visible since the initial distribution of mass tracers has much less symmetries
than in the other cases\footnote{It is important to note here that the reduced symmetry of the initial particle
distribution when using an initial glass or quaquaversal distribution alone did not have an effect on the 
amount of fragmentation observed by \cite{Wang:2007a}.}. Similarly to the TCM case, at the lowest mass resolution, the power
spectrum very slightly underestimates the density at intermediate wave numbers.
\end{itemize}

At late times, when the non-linear part of the power spectrum is dominated by haloes,
the outcome of this comparison is very different. The standard PM results
all converge to the same spectrum, apart from some excess power at the highest wave
numbers in the lowest mass resolution case.
For the new methods, no similar convergence can be observed. As we have already 
discussed above, at densities $\delta\gtrsim 100$ and hence in regions of strong mixing inside of haloes,
the linear approximation to the phase space distribution function breaks down
leading to overestimated densities in the centres of haloes. The TCM power spectra
are the highest, their amplitude at large $k$ decreasing with increasing mass resolution. 
The T4PM spectra are somewhat
lower, but qualitatively still show the same deviation from the standard PM result.
It is obvious that the reduced difference when going from TCM to T4PM is due to 
increased accuracy in modelling the mass distribution of the tetrahedra.
The sudden drop in power when going from $256^3$ to $512^3$ flow tracers 
is however present in both cases and clearly hints at the critical importance to resolve
the distribution function accurately in high density regions. The goal of a future refinement approach thus has
to be to insert the additional flow tracers only where needed.

Finally,  in Figure~\ref{fig:power_sheet}, we show power spectra computed directly
from the dark matter sheet tesselation for standard PM and TCM. Here, instead of
performing a CIC deposit at the particle and mass tracer positions, we projected
the tetrahedra directly into a three dimensional array of $256^3$ cells. We then
used the FFT to compute the power spectra. We see now a resolution dependent
increase in power at large $k$ also in the standard PM case (upper panel). This excess is smaller
than in the TCM case (lower panel) where the overestimated densities in haloes have already
dynamically lead to a run-away increase. This is clear further evidence that it is
indeed the break-down of the piecewiese linear approximation in halos that 
causes the observed excess power at high $k$ compared to the power spectra 
obtained with CIC for the standard PM method. It is thus also clear that 
we cannot quote a maximum value of $k$ up to which the method performs
well in the deeply non-linear regime. Since it fails in small regions of space (the
centers of haloes), these will be spread out in Fourier space over large ranges
of wavenumbers $k$. At the same time, the density PDFs indicate that the 
method provides well converged results up to overdensities of $\delta\lesssim100$.
 

\section{Discussion}
\label{sec:discussion}

Numerical simulations of large scale structure formation are by now a
standard tool in physical cosmology \citep[e.g.][to name but a
few]{Peebles:1971, Davis:1985, Efstathiou:1985, Bertschinger:1998,
  Heitmann:2008, Springel:2005a, Boylan-Kolchin:2009}. Thousands of
studies have ran such calculations or used results from them. They are
used to calibrate analytical approaches~\citep[e.g.][]{Sheth:2001,
  Pueblas:2009}, create Mock galaxy catalogues \citep{Kauffmann:1999,
  Springel:2005a, Conroy:2006}, connect gravitational lensing
measurements to theory \citep{Bartelmann:1998, Jain:2000,
  Schrabback:2010, Becker:2011}, and measure cosmological
parameters~\citep{Seljak:2005, Cole:2005, Viel:2006}. Remarkably, to
the best of our knowledge, the basic methodology and underlying
numerical algorithms, which are practical for three dimensional
calculations, have not evolved. For more than 20 years all algorithms
used today have been known \citep[][and references
therein]{Hockney:1981} and refinements have mostly focused on increasing
the dynamic range of the force calculation.  \cite{Doroshkevich:1980} applied the particle--mesh (PM)
technique to study cosmological structure formation in two dimensions, \cite{Klypin:1983} in
three dimensions. \cite{Efstathiou:1985} discussed the
extension to particle-particle particle-mesh (P$^3$M).
  \cite{Barnes:1986}~developed the
tree--algorithm, \cite{Couchman:1991} gave the first implementation
of AP$^3$M the mesh refined particle-particle particle-mesh
code, and \cite{Xu:1995} combined the tree approach for short-range
and the PM approach for long-range forces. 
These approaches are found in all modern cosmological
simulations codes which are applied to study flow problems of
collisionless dark matter evolving under its own
gravity~\citep[e.g.][]{Bryan:1997, Pearce:1997, Teyssier:2002,
  Wadsley:2004, Springel:2005}.
 
The difference between these algorithms is solely how the gravitational
potential is evaluated on the set of massive particles they seek to
evolve. These particles are both tracers of the flow, and sources of
the gravitational potential. The method which we introduced here is, to
the best of our knowledge, the first approach that conceptually
separates these roles. The potential is sourced by pseudo-particle
approximations to the tetrahedral decomposition of the dark matter sheet
in phase space which is created by the tracers of the fluid flow
(particles). This concept has many advantages and will likely enable a
larger class of methods to be developed in the future. One key aspect
is that the fluid flow tracers track the corners of the fluid volumes one
is considering. The dark matter sheet evolves in phase space in a
volume preserving fashion. For sufficiently resolved situations, all
such methods can reduce the collisionality that has plagued
standard N--body codes for decades. The proposal here is about the
simplest implementation of these ideas one can imagine. Given that it
already gives the benefits we documented one can be optimistic that
even better accuracy can be achieved with further improvements. 

The method proposed explicitly relies on the dark matter sheet being
tracked accurately. However, our results in this paper indicate that 
Lagrangian motion of the flow tracers is insufficient to track the full
evolution of the phase space sheet inside of haloes, where mixing
occurs. The obvious next step is thus to allow for a refinement of the
piecewise linear approximation that we adopted. We have outlined
in Section~\ref{sec:linearerrors} a straightforward way to arrive at
such an adaptive approach that we will discuss in detail in a future
publication, and that logically extends our methods also into 
those regions of the flow where strong mixing occurs without wasting
resolution on those regions that are dominated by larger scale linear 
motion (i.e. translation and shear) where our method is accurate.

As the GPU renderings in Figure~\ref{fig:teaser} demonstrate,
but see also AHK12 and \cite{Kaehler:2012}, it is not difficult to find a highly
accurate dark matter density with high fidelity and dynamic range. One
may further take advantage of that high quality density field when
calculating the forces acting on the fluid instead of relying on pseudo-particle
approximations. A promising area here is
the ubiquitous role of GPU based super-computing. {\sc OpenGL} e.g. allows
to directly render into three dimensional textures (see \citealt{Kaehler:2012},
where we demonstrate specifically how this can be achieved for the
tetrahedra). This will allow to
use the full information of the dark matter sheet projected into
real space to be used as the source for the potential calculation. We
have a already developed a prototype which is already of great
convenience for visualization and likely can be extended also to ``on
the fly`` force calculations.  In particular, fast Fourier methods have already
been ported to the graphics hardware allowing one to keep data on the
device rather than shipping it back and forth between the host and the
graphics card.  The way AMR codes such as {\sc
  Enzo}~\citep{Bryan:1997, Bryan:2001, OShea:2004} or {\sc
  Ramses}~\citep{Teyssier:2002} solve the gravitational potential can
be extended to take advantage of tetrahedral PM.  Such implementations could
remain fully adaptive in space and extend the fixed force resolution
calculations we have presented here.


\section{Summary}
\label{sec:summary}

Following the dark matter sheet in phase space has allowed us to
improve upon traditional particle--mesh simulation codes. Our
rather simple extension shows great promise and already alleviates the
numerical instabilities seen in hot/warm dark matter simulations for
more than two decades.  We summarize our findings as follows:

(1) We discuss a novel method that explicitly makes use of the three
dimensional nature of the phase space sheet to compute forces in
cosmological $N$-body simulations. Using a piecewise linear approximation
to the phase space distribution function, consisting in a decomposition of
the sheet into tetrahedra, we estimate the density field using the cloud-in-cell
deposition at pseudo-particle positions approximating the tetrahedra onto a mesh. 
Specifically, in the TCM approach, we approximate each tetrahedron at monopole
order by a single particle at its centroid position, and in the T4PM approach at
quadrupole order by four particles.
Forces are then computed using the fast Fourier transform as in standard PM codes.

(2) Since the proposed method relies on an accurate tracking of the full
phase space sheet, we observe its limitations when the structure of 
the sheet grows faster than can be tracked by the Lagrangian motion
of the tracer particles. This leads to unphysical behaviour  of our method
in high density regions. We outline how this can be circumvented by
adaptive refinement but defer an analysis to future work and instead
focus on known problems of the traditional $N$-body approach.

(3) We investigate several low dimensional test problems. We find that
collisionality is much reduced in plane wave collapse tests, both axis parallel
and oblique to the initial particle distribution. We find that the proposed 
method approaches the correct solution with increasing force resolution.
This is in contrast to the standard PM method where the phase space structure 
gets destroyed quickly by two-body relaxation effects.

(4) Using a two dimensional test problem, we demonstrate how well our
novel method tracks the caustic structure, while the standard $N$-body 
method suffers from growth of small scale noise and collisionality 
that prevent simultaneous accurate solutions of both the large scale and 
small scale caustic structure.

(5) We considered structure formation in a cosmology with an initial power 
spectrum with truncated small scale power. In contrast to the standard PM
approach, we find that our new method shows no signs of artificial fragmentation 
in filaments. We observe that when the piecewise linear approximation breaks
down inside of haloes, our new method leads to the unphysical behaviour discussed
in point (2). At the same time, we obtain a much smoother force field that much more accurately
reflects the forces in moderate density regions, even at high force and low
mass resolution. We find good convergence up to moderate overdensities
 $\delta \lesssim 100$.
 
While currently this clearly presents a limitation for CDM
 simulations where the collapse fraction is essentially unity, in WDM/HDM simulations
 the method allows to reveal -- without any sign of fragmentation -- the sharp caustic 
 features and pronounced one- and two-dimensional structures that are typical
 for these cosmologies and that are lost in artificial fragmentation when standard
 $N$-body methods are employed.


\section*{Acknowledgements}
O.H. acknowledges support from the Swiss National Science Foundation (SNSF)
through the Ambizione fellowship.
T.A. gratefully acknowledges support by the National Science
foundation through award number AST-0808398 and the LDRD program at
the SLAC National Accelerator Laboratory as well as the Terman
fellowship at Stanford University. We gratefully acknowledge the support of Stuart
Marshall and the SLAC computational team, as well as
the computational resources at SLAC. We thank Yu Lu, Sergei Shandarin, Romain Teyssier
and Alexander Hobbs for comments and suggestions. We thank the
anonymous referee for insightful comments that helped to strengthen
the presentation of this paper.

\label{lastpage}

\end{document}